\documentclass[11pt,a4paper]{article}

\usepackage[margin=1in]{geometry}
\usepackage[T1]{fontenc}
\usepackage[utf8]{inputenc}
\usepackage{lmodern}
\usepackage{amsmath,amssymb,amsfonts,mathtools}
\usepackage{mathrsfs}
\usepackage{amsthm}
\usepackage{bm}
\usepackage{authblk}
\usepackage{cite}
\usepackage{hyperref}
\usepackage[nameinlink,noabbrev]{cleveref}
\usepackage{enumitem}
\usepackage{xcolor}
\hypersetup{colorlinks=true,linkcolor=blue,citecolor=blue,urlcolor=blue}

\newcommand{\F}{\mathcal{F}}
\newcommand{\Gg}{\mathcal{G}}

\newcommand{\GammaVDW}{\Gamma^{\rm VDW}}
\newcommand{\Mp}{M_p}
\newcommand{\dd}{\mathrm{d}}
\newcommand{\Tr}{\mathrm{Tr}}
\DeclareMathOperator{\Det}{Det}
\newcommand{\D}{\mathcal D}
\newcommand{\Diff}{\mathrm{Diff}}

\newcommand{\OO}{\mathcal O}
\newcommand{\MM}{\mathfrak M}

\numberwithin{equation}{section}

\title{Off-shell equivalence in quantum field theory and gravity\\~\\}

\author[1,2]{Iber\^e Kuntz\thanks{\href{mailto:kuntz@fisica.ufpr.br}{kuntz@ufpr.br}}}
\author[2,3,4]{Stefano Liberati\thanks{\href{mailto:liberati@sissa.it}{liberati@sissa.it}}}
\affil[1]{\em Departamento de F\'isica, Universidade Federal do Paran\'a,
Rua Cel. Francisco Her\'aclito dos Santos, 100, 81531-980, Curitiba -- PR, Brazil}
\affil[2]{\em SISSA, via Bonomea 265, 34136, Trieste, Italy}
\affil[3]{\em INFN, Sezione di Trieste, Via Valerio 2, 34127 Trieste, Italy}
\affil[4]{\em IFPU -- Institute for Fundamental Physics of the Universe, Via Beirut 2, 34151 Trieste, Italy}
\date{}

\begin{document}
\maketitle

\begin{abstract}
Field redefinitions connect many formulations of the same physics, but the standard equivalence theorem is an on-shell result and cannot be used to decide when two quantum descriptions are equivalent off shell. This paper develops an operational criterion for that problem in terms of the Vilkovisky--DeWitt effective action. The central idea is that equivalence should be tested on scalar observables built by pairing configuration-space tensors with admissible probes. This makes the criterion sensitive to the observable class under consideration and separates the usual on-shell notion of equivalence from the stronger off-shell notions needed in gravity, cosmology and non-equilibrium quantum field theory. Metric $f(R)$ gravity and a scalar field theory example serve as prototypes, showing how the formal criterion distinguishes local, branchwise and genuinely global equivalence. In particular, we show that metric \(f(R)\) gravity and its auxiliary-field reformulation yield the same quantum theory when the auxiliary constraint is enforced in the path integral. This is distinct from quantizing the corresponding scalar--tensor action with the metric and scalar treated as independent integration variables, which defines a different quantum theory. Apparent quantum inequivalences can then be traced to comparisons between different quantum objects, rather than to a failure of actual equivalences. This leads to general and precise notions of local and global equivalence under both field redefinitions and auxiliary-variable extensions.
\end{abstract}

\tableofcontents

\section{Introduction}
\label{sec:introduction}

Field redefinitions are ubiquitous in quantum field theory. They are commonly used to turn a higher-derivative or non-minimally coupled theory into a form that is easier to quantize or interpret. In effective field theory (EFT), for example, they are used to remove redundant operators and to choose convenient bases for simplifying calculations of amplitudes \cite{Georgi:1991ch,Arzt:1993gz,Criado:2018sdb}. Modern geometric treatments of EFT take this further, turning field redefinitions into coordinate changes on the relevant field manifold \cite{Alonso:2015fsp,Cohen:2022uuw,Cohen:2023wxr,Cohen:2024geometry,Cohen:2025nonlocal,Cohen:2025building}. In gravity and cosmology field redefinitions often appear as conformal rescalings and frame transformations, which are standard tools in scalar--tensor theories, $f(R)$ and higher-derivative models \cite{Starobinsky:1980te,Whitt:1984pd,Sotiriou:2008rp,DeFelice:2010aj}.

In particle physics the use of field redefinitions is justified by invoking the equivalence theorem \cite{Chisholm:1961tha,Kamefuchi:1961sb,Rebhan:1987zm}, which guarantees that local, perturbative and invertible changes of field variables do not change the $S$-matrix. This result explains why one can move between different off-shell Lagrangian representatives while keeping scattering data unchanged. It also underlies the familiar EFT statement that operators proportional to the equations of motion are redundant for on-shell amplitudes \cite{Georgi:1991ch,Arzt:1993gz,Criado:2018sdb}.

There is, however, a basic limitation in the equivalence theorem. Because it is an on-shell statement, it tells us what happens to scattering amplitudes, but it does not provide a general criterion for the equivalence of off-shell quantum theories. This limitation is often harmless in collider physics, where asymptotic particle states are the natural observables, but it becomes an issue once one moves away from that setting. In gravity, cosmology, non-equilibrium dynamics and many other fields, the relevant observables are not exhausted by $S$-matrix elements. In fact, many questions of physical interest are questions about dynamics rather than scattering. In these cases, one is often interested in expectation values, response functions, in-in correlators and other dynamical observables \cite{Schwinger:1960qe,Keldysh:1964ud,Kubo:1957mj}. These are off-shell quantities. In the standard formulation of QFT they are not invariant under general field redefinitions.

Gauge theories and gravity make this issue particularly sharp as the usual effective action is not only sensitive to the parametrization of the fields off shell, it is also gauge dependent. In such situations one needs a formulation in which the off-shell quantum action has the correct geometrical transformation properties. This is precisely the role of the Vilkovisky--DeWitt construction \cite{Vilkovisky:1984st,DeWitt:1987yk,Barvinsky:1985an,Rebhan:1987nw,Finn:2019aip}. The Vilkovisky--DeWitt (VDW) effective action is designed to turn the effective action into a scalar on configuration space, and in gauge theories to remove the spurious gauge-condition dependence of the ordinary construction.

This point is closely tied to a debate that has reappeared many times in the gravity and cosmology literature, namely the relation between the Jordan and Einstein frames. At the classical level the map between the two descriptions is usually presented as a field redefinition, more precisely as a combination of a Weyl rescaling and a scalar-field reparametrization. This has led to the question of whether one frame is more physical than the other. The literature contains many formulations of this problem. Some authors emphasize mathematical equivalence and the convenience of choosing whichever frame is easier for the calculation. Others stress that physical interpretation depends on what is held fixed, how matter is coupled, what observables are considered, and whether the comparison is made on shell or off shell. Despite the ongoing discussion, it is worth noting that the classical action is a scalar on configuration space, so its first functional derivatives transform covariantly and there is a local one-to-one correspondence between solutions in the two frames. The real issue takes place at the quantum level, for the naive effective action does not feature frame covariance \cite{Dicke:1961gz,Wagoner:1970vr,Magnano:1993bd,Faraoni:2006fx,Postma:2014vaa,Kamenshchik:2014waa,Ohta:2017fqs,Ruf:2017xon,Falls:2018utl,Karamitsos:2017elm,Finn:2019aip,Casadio:2022ndh}.

The point of the present paper is to formulate criteria of equivalence for classical and quantum field theories. The basic criterion is that the same observables, in a given class, should agree. Our notion of equivalence is not absolute, but rather relative to the observable class of interest (such as scattering amplitudes for collider physics). This allows the generalization of the equivalence theorem to dynamical settings. The VDW effective action\footnote{In this paper we formulate the quantum criterion using the standard in--out/Euclidean VDW effective action. Genuine real-time response and in--in observables require the corresponding Schwinger--Keldysh/closed-time-path version of the geometric effective action. We do not develop that extension here.}
will serve as a proxy to off-shell observables. We use \(f(R)\) gravity and the Jordan/Einstein-frame map as prototypes, and we use a simple scalar field example to separate on-shell equivalence from equality of off-shell response functions. These examples force us to distinguish three notions of equivalence: (i) equivalence under field redefinitions and auxiliary-variable extensions, (ii) local, branchwise and global equivalence, and (iii) classical and quantum equivalence.

A central theme of the paper is that equivalence is only meaningful relative to a class of
observables. In particle physics one may decide that the physically relevant probes are just those
that extract on-shell amplitudes. Then two descriptions may be fully equivalent for that restricted
operational class even if their off-shell correlation functions differ. In semiclassical gravity or non-equilibrium physics one usually needs a larger class of probes, because the relevant observables are response
functions, expectation values and dynamical correlators. This suggests a useful distinction between
a mathematically separating\footnote{Here ``separating'' means that the probe class is rich enough to distinguish the theory tensors under consideration. Namely, if two such tensors have the same pairing with every probe in the class, then they are identified.} probe space and a physically admissible one. The first is large enough to
separate the theory tensors. The second encodes the observables one regards
as physical or accessible in a given experiment. This distinction will let us phrase equivalence in a way that is both
general and operational.

Our formalism provides a unified and precise language for several questions that are often phrased too loosely and without proper qualification in the literature, such as whether two actions describe the same theory or whether different frames are physically equivalent. It also separates genuine failures of equivalence from comparisons in which different quantum objects have been used, as in the quantization of $f(R)$ and its different forms. Moreover, we believe its usefulness goes well beyond the formal problem of frame equivalence. It bears directly on redundant operators and basis changes in non-equilibrium and gravitational EFTs.

The paper is organized as follows. In Section~\ref{sec:fr_seed} we begin with the \(f(R)\) example and use it to expose the distinction between direct reparametrizations, auxiliary-variable extensions and local domains of validity. In Section~\ref{sec:quantum_warmup_section} we use a scalar example to separate the on-shell content of the equivalence theorem from the behavior of off-shell response functions. In Section~\ref{sec:vdw_section} we review the Vilkovisky--DeWitt effective action, which will provide the off-shell quantum object used in the comparison. In Section~\ref{sec:observables_section} we introduce observables as pairings of theory tensors with probes, and we explain the role of different probe classes. In Section~\ref{sec:classical_equiv_section} we formulate classical equivalence, including direct and parent-mediated equivalence and the distinction between local, branchwise and global domains. Section~\ref{sec:quantum_equiv_section} gives the corresponding quantum criterion in terms of renormalized VDW effective actions. Finally, in Section~\ref{sec:examples_section} we exemplify our formalism in a simple scalar EFT and discuss the equivalence of \(f(R)\) to its different representations at the quantum level.

\section{A classical warm-up: $f(R)$ and frames}
\label{sec:fr_seed}

\subsection{$f(R)$ and its scalar--tensor form}

Metric $f(R)$ gravity provides a useful starting point for formulating our criteria of equivalence \cite{Starobinsky:1980te,Whitt:1984pd,Sotiriou:2008rp,DeFelice:2010aj}. It is simple enough that the transformed theory can be displayed explicitly, but nontrivial enough to involve the main structures needed later.

Consider the metric theory
\begin{equation}
S_f[g]
=
\frac{\Mp^2}{2}\int \dd^4x\,\sqrt{-g}\,f(R)\, ,
\label{eq:Sf_intro_section}
\end{equation}
where $M_p$ is the Planck mass.
The configuration space here is the space $\F_g$ of Lorentzian metrics on
spacetime, modulo the usual subtleties about signature and regularity that we leave implicit. The
gauge group is $\Diff(M)$, namely the group of diffeomorphisms on the spacetime $M$.

A standard way of lowering the derivative order is to introduce an auxiliary scalar $\chi$ \cite{Whitt:1984pd,Sotiriou:2008rp,DeFelice:2010aj}. The
cleanest way to do this is first to impose $\chi=R[g]$ with a Lagrange multiplier field $\lambda(x)$:
\begin{equation}
S_\lambda[g,\chi,\lambda]
=
\frac{\Mp^2}{2}\int \dd^4x\,\sqrt{-g}\,
\Big(
f(\chi)+\lambda\big(R-\chi\big)
\Big)\, .
\label{eq:Slambda_intro_section}
\end{equation}
This action is defined on the enlarged configuration space
\begin{equation}
\F_{\rm ext}=
\F_g\times \F_\chi\times \F_\lambda\, ,
\label{eq:Fext}
\end{equation}
where $\F_\chi$ and $\F_\lambda$ denote the spaces of scalar fields used for the auxiliary variable
and the Lagrange multiplier, respectively.

The equations of motion for $\lambda$ and $\chi$ are
\begin{align}
\label{eq:chisol1}
\frac{\delta S_\lambda}{\delta \lambda}
&= \frac{\Mp^2}{2}\sqrt{-g} (R - \chi) = 0
&&\Longrightarrow
&\chi&=R
&&\text{(everywhere)},
\\
\frac{\delta S_\lambda}{\delta \chi}
&= \frac{\Mp^2}{2}\sqrt{-g} (f'(\chi) - \lambda) = 0
&&\Longrightarrow
&\lambda&=f'(\chi)
&&\text{(everywhere)} .
\end{align}
Note that both solutions are obtained without any domain restrictions and are thus valid everywhere in $\mathcal F_{\rm ext}$. Eliminating $\chi$ by evaluating $S_\lambda$ at the $\chi$ equation of motion recovers Eq.~\eqref{eq:Sf_intro_section}:
\begin{equation}
S_\lambda[g,\chi=R,\lambda] = S_f[g]
\qquad \text{(everywhere)}
,
\label{eq:equiv1}
\end{equation}
for all $\lambda$.
On the other hand, eliminating $\lambda$ instead gives the two-field action most commonly found in the literature:
\begin{equation}
\widehat S[g,\chi]
\equiv
S_\lambda[g,\chi,\lambda=f'(\chi)]
=
\frac{\Mp^2}{2}\int \dd^4x\,\sqrt{-g}\,
\Big(
f(\chi)+f'(\chi)\big(R-\chi\big)
\Big)
\ ,
\label{eq:Shat_intro_section}
\end{equation}
which lives on $\widehat{\F}=\F_g\times \F_\chi$. It is important to distinguish between the theory $(\widehat S, \chi=R)$, which inherits all data from $S_\lambda$, and the standalone theory given solely by the action $\widehat S$.
Indeed, differentiating Eq.~\eqref{eq:Shat_intro_section} with respect to $\chi$ gives:
\begin{equation}
\frac{\delta \widehat S}{\delta \chi}
=
\frac{\delta S_\lambda}{\delta \chi}\bigg|_{\lambda=f'(\chi)}
+
\frac{\delta S_\lambda}{\delta \lambda}\bigg|_{\lambda=f'(\chi)}
\frac{\delta \lambda}{\delta \chi}
\ .
\label{eq:chi_eom_intro_section}
\end{equation}
Thus the equation of motion for $\chi$ in the theory $(\widehat S, \chi=R)$ reads:
\begin{equation}
\frac{\Mp^2}{2}\sqrt{-g}\,f''(\chi)\,(R-\chi)
=0
\ ,
\label{eq:chisol2}
\end{equation}
which is trivially true from Eq.~\eqref{eq:chisol1}.

However, had we started out from $\widehat S$ as the usual starting point in the literature, without knowing about $S_\lambda$ (and hence about Eq.~\eqref{eq:chisol1} beforehand), Eq.~\eqref{eq:chisol2} would no longer be trivial. In the standalone theory, we would then find the same algebraic solution but with a restricted domain:
\begin{equation}
\chi=R
\qquad
\text{whenever } f''(\chi)\neq 0\, .
\end{equation}
Substituting this solution back into Eq.~\eqref{eq:Shat_intro_section} also gives
Eq.~\eqref{eq:Sf_intro_section}:
\begin{equation}
\widehat S[g,\chi=R] = S_f[g] \ , \qquad \text{for } f''(\chi)\neq 0
\ .
\end{equation}
Therefore, the metric $f(R)$ action is recovered from the two-field action $\widehat S$, but this time the equivalence holds only on the domain where $f''(\chi)\neq 0$. This is to be contrasted with Eqs.~\eqref{eq:chisol1} and \eqref{eq:equiv1}, which are valid everywhere in configuration space. Such difference can be understood from Eq.~\eqref{eq:chi_eom_intro_section}. A solution for $\chi$ in the theory $S_\lambda$ is necessarily a solution in $\widehat S$, but the opposite is not true. In fact, the theory $\widehat S$ contains less information than $S_\lambda$ since the former was obtained by evaluating the latter over the specific direction $\lambda=f'(\chi)$. So if one starts off with $S_\lambda$, the solution $\chi=R$ holds globally, thus $S_f, S_\lambda$ and $(\widehat S, \chi=R)$ are globally equivalent. On the other hand, starting from the standalone $\widehat S$, the theories $S_f$ and $\widehat S$ are only locally equivalent.

The above discussion also illustrates another point. The relation between the actions above (e.g. Eqs.~\eqref{eq:Sf_intro_section} and
\eqref{eq:Shat_intro_section}) is not an ordinary field redefinition on a fixed field space. The field
content has changed. What relates the two descriptions is an extension of the configuration space (see Eqs.~\eqref{eq:Slambda_intro_section} and \eqref{eq:Fext})
followed by a reduction (Eqs.~\eqref{eq:equiv1} and \eqref{eq:Shat_intro_section}). This simple observation is often hidden in the standard presentation, but it
will matter later when we discuss quantum equivalence.

\subsection{Jordan and Einstein frames}

The next step is the passage to the scalar--tensor Jordan frame \cite{Whitt:1984pd,Magnano:1993bd,Sotiriou:2008rp,DeFelice:2010aj}. Define
\begin{equation}
\phi=f'(\chi)\, .
\label{eq:phi_legendre_intro_section}
\end{equation}
Whenever $f''(\chi)\neq 0$, this relation can be inverted locally to give $\chi=\chi(\phi)$. On the
corresponding domain in $\widehat{\F}$, the map $(g,\chi)\mapsto(g,\phi)$ is a genuine local field
redefinition. In terms of the new scalar, the action becomes
\begin{equation}
S_J[g,\phi]
=
\frac{\Mp^2}{2}\int \dd^4x\,\sqrt{-g}\,
\Big(
\phi\,R[g]-U(\phi)
\Big)\, ,
\qquad
U(\phi)=\phi\,\chi(\phi)-f(\chi(\phi))\, .
\label{eq:SJ_intro_section}
\end{equation}

This step has a different character from the previous one. No variable is being eliminated and no
equation of motion is being solved. One is simply changing coordinates on the enlarged field
space. The map from $\widehat S[g,\chi]$ to $S_J[g,\phi]$ is an ordinary local
reparametrization in $\widehat{\mathcal F}$.

From the Jordan frame one then reaches the Einstein frame by another local field redefinition \cite{Dicke:1961gz,Wagoner:1970vr,Magnano:1993bd,Faraoni:2006fx}:
\begin{equation}
\hat g_{\mu\nu}\equiv\phi\, g_{\mu\nu},
\qquad
\sigma\equiv\sqrt{\frac{3}{2}}\,\Mp\ln\phi\, ,
\label{eq:Einstein_map_intro_section}
\end{equation}
which is well defined on the domain $\phi>0$. In these variables the action takes the familiar form
\begin{equation}
S_E[\hat g,\sigma]
=
\int \dd^4x\,\sqrt{-\hat g}\,
\left[
\frac{\Mp^2}{2}\hat R
-\frac{1}{2}(\hat\nabla\sigma)^2
-V_E(\sigma)
\right],
\label{eq:SE_intro_section}
\end{equation}
with
\begin{equation}
V_E(\sigma)=\frac{\Mp^2}{2}\,\frac{U(\phi(\sigma))}{\phi(\sigma)^2}\, .
\end{equation}

This is the place where the standard frame debate usually starts \cite{Postma:2014vaa,Kamenshchik:2014waa,Ohta:2017fqs,Ruf:2017xon,Falls:2018utl,Finn:2019aip,Casadio:2022ndh}. Different answers in the literature
usually reflect different implicit choices about observables and about the level at which the
comparison is being made.
For example, the proper time along a worldline $x^\mu(\lambda)$ in the Jordan frame is:
\begin{equation}
\tau_J[g]
=
\int d\lambda\,
\sqrt{
-g_{\mu\nu}\dot x^\mu\dot x^\nu
}\, .
\label{eq:tau_J_def}
\end{equation}
If one then naively defines the same-looking Einstein-frame expression
\begin{equation}
\tau^{\rm naive}_E[\hat g]
=
\int d\lambda\,
\sqrt{
-\hat g_{\mu\nu}\dot x^\mu\dot x^\nu
}\, ,
\label{eq:tau_E_same_looking}
\end{equation}
one has chosen a different clock. Then the measurements in each frame would indeed differ:
\begin{equation}
    \tau_J[g]
    \neq
    \tau^{\rm naive}_E[\hat g]
    \ .
\end{equation}
This is, however, a different experiment, not a different prediction
for the same observable.
The proper time in the Einstein frame is correctly given by the pullback of Jordan's:
\begin{equation}
\tau_E[\hat g,\sigma]
=
\int d\lambda\,
\phi(\sigma)^{-1/2}
\sqrt{
-\hat g_{\mu\nu}\dot x^\mu\dot x^\nu
}\, .
\label{eq:tau_J_in_E_variables}
\end{equation}
Thus the proper time is a scalar under the field redefinitions:
\begin{equation}
    \tau_E[\hat g,\sigma]
    =
    \tau_J[g(\hat g, \sigma)]
    \ .
\end{equation}
Its functional form changes, as it should, but its value is the same in the two descriptions.

Therefore, at the classical level the right statement is clear. On the domain where the
transformations are invertible, the Jordan and Einstein descriptions are related by an ordinary field
redefinition. The classical action is a configuration-space scalar:
\begin{equation}
S_E[\hat g, \sigma]
=
S_J[g(\hat g, \sigma), \phi(\sigma)]
\ ,
\end{equation}
thus the corresponding equations of motion transform covariantly:
\begin{equation}
\frac{\delta S_E}{\delta\tilde\varphi^A}
=
\frac{\delta\varphi^B}{\delta\tilde\varphi^A}
\frac{\delta S_J}{\delta\varphi^B}
\ ,
\end{equation}
where $\varphi^A = (g, \phi)$ and $\tilde\varphi^A=(\hat g, \sigma)$. As long as the redefinition $\tilde\varphi(\varphi)$ is locally invertible, there is a one-to-one correspondence between solutions in the two frames.
Therefore, they are locally equivalent as classical theories. At this stage there is no
reason to single out one frame as fundamentally physical. What matters is that physical quantities
must be computed consistently in whichever variables one uses.

The subtlety is that this equivalence is local, not global. The Einstein-frame map
requires $\phi>0$. The Jordan-frame scalar itself was introduced through $\phi=f'(\chi)$, which
required $f''(\chi)\neq 0$ to be locally invertible. The reduction back to the metric $f(R)$ form also
required $f''(\chi)\neq 0$. Thus the chain of transformations from metric $f(R)$ to Jordan frame to Einstein frame is
naturally a chain of local statements on local domains. Any general notion of equivalence should
therefore be able to talk about equivalence on a field-space patch, not only about global
equivalence on the full configuration space.

\subsection{Lessons for classical equivalence}
\label{sec:lessons_classical_equivalence}

The \(f(R)\) chain already teaches several lessons that shall be abstracted below.
The first lesson is that equivalence need not be a direct map between endpoint
field spaces. The passage from metric \(f(R)\) gravity to the scalar--tensor
description is not an ordinary change of variables on the space of metrics. One
first enlarges the configuration space by introducing auxiliary variables, and
only then obtains the endpoint descriptions by reduction and reparametrization.
Thus a useful notion of equivalence must allow for parent constructions, not
only for direct field redefinitions.

Secondly, the notion of an auxiliary field is endpoint
dependent. In the metric \(f(R)\) endpoint, the scalar introduced in the parent
description is eliminated by its equation of motion. In the scalar--tensor
endpoint, the corresponding scalar is part of the physical field content. The
same parent variable may therefore be auxiliary relative to one endpoint and
non-auxiliary relative to another. This distinction will matter even more after
quantization, because it determines which variables are sourced and which
variables are integrated out as part of a reduction.

The third lesson is locality in configuration space. The reduction from the
two-field form to the metric \(f(R)\) action requires a branch on which
\(f''\neq0\). The Jordan variable \(\phi=f'(\chi)\) is also only locally
invertible on such a branch. The Einstein-frame map further requires
\(\phi>0\). Hence the classical relation between the metric, Jordan and
Einstein descriptions is not a single unrestricted equivalence on the full
configuration space. It is a chain of local equivalences on domains where the
corresponding reductions and reparametrizations are well defined.

Lastly, classical field redefinitions are already
geometrical. On the domain where the Jordan--Einstein map is invertible, the
classical action is a scalar functional on configuration space. The equations of
motion then transform covariantly, and solutions are mapped into solutions. At
the classical level, this is the appropriate meaning of frame equivalence. There
is no need to declare one frame more fundamental, provided one compares the same
physical quantities and stays within the domain of the map.

These lessons motivate the classical part of the formalism developed below. A
classical equivalence criterion must distinguish direct reparametrizations from
parent-mediated equivalences, must keep track of endpoint-dependent reductions,
and must be formulated on field-space domains rather than only globally. These criteria are not enough for addressing quantum equivalence. For that one must specify the
measure, the source prescription, the field-space metric, and the effective
quantum object being compared.

\section{A quantum warm-up: scattering amplitudes and response functions}
\label{sec:quantum_warmup_section}

\subsection{On-shell observables and the equivalence theorem}
\label{sec:quantum_scalar_source_diagnostic}

In high-energy EFT, field redefinitions are usually used perturbatively, close to the identity, to remove
redundant operators in the bare action and simplify the operator basis \cite{Georgi:1991ch,Arzt:1993gz,Criado:2018sdb}. In this context, the equivalence theorem guarantees that the \(S\)-matrix is unchanged. This equivalence, however, holds only on-shell, hence it cannot be used as a tool for off-shell observables, such as response functions.

We shall now separate these two statements in a simple example. We start from a free
scalar theory and rewrite it in nonlinear field coordinates. Any interaction
generated in this way is necessarily a coordinate artifact. The example is useful
because it shows, in the same model, how the equivalence theorem works for Lehmann--Symanzik--Zimmermann (LSZ)
observables and why the same argument does not apply to generic off-shell response
functions.

Consider a real scalar field with quadratic action
\begin{equation}
S_0[\phi]
=
\frac12\int d^dx\,\phi K\phi ,
\label{eq:free_scalar_action_probe}
\end{equation}
where \(K\) is the inverse propagator. In Lorentzian signature one may take
\(K=-(\Box+m^2)\), with the usual \(i0\) prescription. In the Euclidean
functional integrals below, \(K\) denotes the corresponding positive elliptic
operator. We write
\begin{equation}
\Delta=K^{-1}
\label{eq:prop}
\end{equation}
for the propagator. The theory is free, so connected ordinary \(n\)-point
functions of the field \(\phi\) vanish for \(n>2\).

Now perform the local field redefinition
\begin{equation}
\phi(x)=F(\psi(x))
=
\psi(x)+\frac{a}{M^2}\psi(x)^3,
\qquad a>0 ,
\label{eq:cubic_field_redefinition_probe}
\end{equation}
where $M$ is an arbitrary mass scale and $a$ a dimensionless parameter.
For \(a>0\), this is a pointwise global diffeomorphism, since
\begin{equation}
F'(\psi)=1+\frac{3a}{M^2}\psi^2>0 .
\end{equation}
We use a cubic rather than a quadratic redefinition because it produces a
first-order correction to the two-point susceptibility below. The price is that
the redundant interaction first appears as a quartic rather than a cubic vertex.

Substituting \eqref{eq:cubic_field_redefinition_probe} into
\eqref{eq:free_scalar_action_probe} gives
\begin{equation}
S_F[\psi]
\equiv
S_0[F(\psi)]
=
\frac12\int d^dx\,\psi K\psi
+
\frac{a}{M^2}\int d^dx\,\psi^3 K\psi
+
O(M^{-4}) .
\label{eq:free_scalar_redefined_action_probe}
\end{equation}
Thus a free theory now appears to contain an interaction. In momentum space,
suppressing the momentum-conserving delta function, the ordinary four-point 1PI
vertex is proportional to
\begin{equation}
\Gamma^{(4)}_{\rm ord}(p_1,p_2,p_3,p_4)
=
\frac{a}{M^2}
\bigl[
K(p_1)+K(p_2)+K(p_3)+K(p_4)
\bigr]
+
O(M^{-4}) ,
\label{eq:ordinary_four_vertex_probe}
\end{equation}
up to the conventional overall combinatorial normalization. This vertex is nonzero off shell. In the LSZ reduction formula, inverse-propagator
factors amputate the external legs of the connected four-point function. At the
order considered here, this leaves \(\Gamma^{(4)}_{\rm ord}\), up to external-leg
residues, evaluated at \(K(p_i)=0\) \cite{Lehmann:1954rq}. Since every term in
Eq.~\eqref{eq:ordinary_four_vertex_probe} contains the inverse propagator of an
external leg,
\begin{equation}
\mathcal A^{(4)}_{\psi}
\propto
\left.
\Gamma^{(4)}_{\rm ord}(p_1,p_2,p_3,p_4)
\right|_{K(p_i)=0}
=
0 .
\label{eq:lsz_zero_probe}
\end{equation}
The off-shell vertex has changed, but the on-shell amplitude has not. This is
the elementary content of the equivalence theorem in this example
\cite{Lehmann:1954rq,Chisholm:1961tha,Kamefuchi:1961sb,Georgi:1991ch,Arzt:1993gz,Criado:2018sdb}.
The fields \(\psi\) and \(F(\psi)\) are different off-shell operators, but they
interpolate the same one-particle state perturbatively. LSZ probes are
insensitive to this difference.

The path integral makes the distinction sharper. Starting from
\begin{equation}
Z[J]
=
\int D\phi\,
\exp\left\{
iS_0[\phi]+iJ\cdot\phi
\right\},
\label{eq:path_integral_phi_probe}
\end{equation}
the change of variables \(\phi=F(\psi)\) gives
\begin{equation}
Z[J]
=
\int D\psi\,
\det\!\left(\frac{\delta F}{\delta\psi}\right)
\exp\left\{
iS_F[\psi]+iJ\cdot F(\psi)
\right\}.
\label{eq:path_integral_exact_rewrite_probe}
\end{equation}
Note that this is just a change of integration variables, hence Eqs.~\eqref{eq:path_integral_phi_probe} and \eqref{eq:path_integral_exact_rewrite_probe} are trivially equivalent. In particular, the source
has not become a source for \(\psi\). It has become a source for the composite
operator \(F(\psi)\). Therefore the exact off-shell identity is
\begin{equation}
\bigl\langle
\phi(x_1)\cdots\phi(x_n)
\bigr\rangle_{S_0}
=
\bigl\langle
F(\psi)(x_1)\cdots F(\psi)(x_n)
\bigr\rangle_{S_F}.
\label{eq:composite_correlator_equivalence_probe}
\end{equation}
This is not the statement that ordinary correlators of \(\psi\) equal ordinary
correlators of \(\phi\). Nor is this the content of the equivalence theorem.

The equivalence theorem addresses a different question. For the sake of the argument, suppose that Alice and Bob were handed the classical theories $S_0[\phi]$ and $S_F[\psi]$, respectively, without knowing their origin, and asked to assess whether the corresponding quantum theories were equivalent. To quantize it, Alice would proceed with Eq.~\eqref{eq:path_integral_phi_probe} as usual. Bob, without knowing better, would also follow the standard practice and write down the functional generator with linear coupling to the source:
\begin{equation}
Z_{\rm lin}[J]
=
\int D\psi\,
\exp\left\{
iS_F[\psi]+iJ\cdot\psi
\right\}.
\label{eq:path_integral_linear_source_probe}
\end{equation}
The generator $Z_{\rm lin}[J]$ is obviously different from $Z[J]$, so they would conclude (correctly but without proper qualification) that the theories are different. However, after computing scattering amplitudes for different processes, they would then realize the S-matrix in both theories is the same and would rush to conclude the theories are actually equivalent. This is the equivalence theorem in play.

The equivalence theorem states that scattering amplitudes computed from $Z_{\rm lin}[J]$ and $Z[J]$ agree. This allows one to forget about the transformations of the source coupling and the measure\footnote{In dimensional regularization, the Jacobian of a
local perturbative redefinition is also usually harmless in EFT applications,
since it contributes local ghost loops with polynomial momentum integrals
\cite{tHooft:1973wag,Criado:2018sdb}.}, and pretend that only the classical action transforms. In other words, ``transform then quantize'' commutes with ``quantize then transform'' for on-shell observables. The latter qualification is very important. Because the equivalence theorem requires the LSZ formula, it only holds for on-shell quantities. The apparent interaction in \eqref{eq:free_scalar_redefined_action_probe} is thus redundant for particle physics: it is not absent as an off-shell vertex, but it
is invisible in scattering experiments.

\subsection{Response functions}

Even in high-energy physics, the S-matrix does not exhaust all the relevant observables. In strongly-interacting systems, such as the quark-gluon plasma in QCD, one is often interested in response functions \cite{McLerran:1984ay,Weldon:1990iw,Jeon:1995zm}. Thus, we shall now turn the same example into a response calculation. For simplicity, we shall consider the susceptibility, which is
the linear response of an expectation value to a weak external source. This captures the same source
logic that appears in more physical response problems. The important fact we want to emphasize is that such response functions are observable quantities in experiments designed to probe dynamics. They measure how a system reacts when a source is varied.

To avoid the complications of the Keldysh-Schwinger formalism \cite{Schwinger:1960qe,Keldysh:1964ud}, which is beyond the scope of this paper, we shall now use the Euclidean path integral. To compute the susceptibility, Alice would start off from the same construction as before, now with Euclidean boundary conditions:
\begin{equation}
\mathcal Z[J]
=
\int \mathcal D\phi\,
\exp\left[
-\frac1\hbar
\left(
S_0[\phi]-\int d^dx\,J(x)\phi(x)
\right)
\right]
\ .
\label{eq:euclidean_source_phi_probe}
\end{equation}
Defining the connected generator $\mathcal W[J]=\hbar\ln \mathcal Z[J]$, the mean field is given by
\begin{equation}
\langle \phi(x)\rangle_J
=
\frac{\delta \mathcal W}{\delta J(x)} ,
\end{equation}
and she finds the susceptibility kernel:
\begin{equation}
\chi_\phi(x,y)
\equiv
\left.
\frac{\delta\langle \phi(x)\rangle_J}{\delta J(y)}
\right|_{J=0}
=
\Delta(x,y),
\label{eq:chi_phi_euclidean_probe}
\end{equation}
where $\Delta(x,y)$ denotes the free propagator \eqref{eq:prop} in position space.

Bob, on the other hand, would compute the ordinary coordinate-linear prescription in the \(\psi\)-coordinate:
\begin{equation}
\mathcal Z_{\rm lin}[J]
=
\int \mathcal D\psi\,
\exp\left[
-\frac1\hbar
\left(
S_F[\psi]
-
\int d^dx\,J(x)\psi(x)
\right)
\right]
\ .
\label{eq:ZY_coordinate_linear_source_probe}
\end{equation}
Bob's susceptibility is:
\begin{equation}
\chi_\psi(x,y)
=
\left.
\frac{\delta\langle\psi(x)\rangle_J}{\delta J(y)}
\right|_{J=0}
=
\frac1\hbar
\langle\psi(x)\psi(y)\rangle_{c,S_F}.
\label{eq:bob_susceptibility_def}
\end{equation}
Expanding perturbatively in $\alpha\equiv a/M^2$, it is straightforward to show that
\begin{equation}
\langle\psi(x)\psi(y)\rangle_{c,S_F}
=
\hbar\Delta(x,y)
-
3\alpha\hbar^2
\bigl[
\Delta(x,x)+\Delta(y,y)
\bigr]\Delta(x,y)
+
O(\alpha^2)
\ ,
\label{eq:bob_two_point_final}
\end{equation}
where a UV regulator is understood for the coincident
propagators.
Since Alice found \(\chi_\phi(x,y)=\Delta(x,y)\), they obtain the relation between their susceptibility:
\begin{equation}
\chi_\psi(x,y)
=
\chi_\phi(x,y)
-
\frac{3a\hbar}{M^2}
\bigl[
\chi_\phi(x,x)+\chi_\phi(y,y)
\bigr]\chi_\phi(x,y)
+
O(M^{-4})
\ .
\label{eq:chi_psi_vs_chi_phi_direct}
\end{equation}
Thus Alice and Bob, working independently and coupling sources linearly to
their own field variables, do not obtain the same susceptibility.

Despite the simplicity of this example, the same logic also applies to response problems of physical interest. In thermal field
theory, an external electromagnetic field \(A_\mu\) defines a source functional
whose second derivatives give the real-time current-current response. The
corresponding spectral density controls photon and dilepton emission from a
quark-gluon plasma \cite{McLerran:1984ay,Weldon:1990iw}. Under a field
redefinition of the microscopic fields, the same electromagnetic perturbation
couples to the transformed current, not to the same-looking current written in
the new variables. Similarly, a metric perturbation \(h_{xy}\) sources the
stress tensor, and the retarded \(T^{xy}T^{xy}\) correlator determines the shear
viscosity through the Kubo formula \cite{Kubo:1957mj,Jeon:1995zm}. These
examples show that the source-level distinction outlined above is not merely a
peculiarity of a free scalar model. It is the general structure of response
observables.

\subsection{Lessons for quantum equivalence}
\label{sec:lessons_quantum_equivalence}

The scalar example complements the \(f(R)\) discussion by isolating the quantum
role of observables. It shows that field-redefinition equivalence is not an
absolute statement about a pair of actions. It is a statement about the class of
observables used to compare the two descriptions. Equivalence is always relative to such a chosen class.

For on-shell scattering observables in a theory with asymptotic
particle states, the standard equivalence theorem is often
sufficient. Local perturbative field
redefinitions do not change the \(S\)-matrix extracted by LSZ. In this sense, operators
generated by the redefinition are redundant relative to on-shell scattering
observables. This is exactly the use of field redefinitions in EFT.

This perturbative on-shell statement should not be confused with a general
non-perturbative criterion for equivalence. Standard perturbative equivalence-theorem arguments do not automatically apply
to finite field-space maps such as the Jordan--Einstein transformation unless
the relevant hypotheses, including the identification of asymptotic fields and
interpolating operators, are checked.

Secondly, equivalence only holds off-shell when one transforms the source coupling accordingly. This is, however, a trivial tautology and of little use because the generating functionals do not depend on the integration variable. Moreover, this requires the knowledge of the field redefinition beforehand. The real problem of equivalence regards the comparison of quantum observables before knowing this map. In this case, as illustrated by Alice and Bob state of affairs, one would naively compute the generating functional by
independently choosing linear sources in each parametrization. This does not represent the same response experiment in a new configuration-space coordinate system as the source is coupled to a different scalar function. It is therefore not surprising that off-shell quantities differ. Since fields are coordinates in configuration space, the linear source coupling is coordinate dependent and therefore has no geometrical meaning. The VDW formalism provides the geometrical framework in which this problem can be formulated covariantly.

\section{The Vilkovisky--DeWitt effective action}
\label{sec:vdw_section}

The previous section showed that the usual equivalence theorem is not a general
criterion for quantum equivalence. It is a perturbative on-shell statement. The theorem
explains why local field redefinitions do not change scattering amplitudes, but
it does not provide an intrinsic off-shell quantum object that can be compared
between parametrizations.

In this section, we shall review the Vilkovisky--DeWitt construction. It provides an off-shell effective action with the correct geometrical transformation law under field redefinitions \cite{Vilkovisky:1984st,DeWitt:1987yk,Barvinsky:1985an,Rebhan:1987nw,Finn:2019aip}. This is the quantum object we will use later when formulating off-shell equivalence. Throughout this paper, we shall adopt the standard in--out/Euclidean version of the effective action. Its extension to real-time in--in observables requires a geometric Schwinger--Keldysh/closed-time-path construction, which lies beyond our scope.

\subsection{Configuration-space covariance}

The basic idea is to treat field configurations as points of a manifold. A field
redefinition is then a change of coordinates on configuration space. We denote
this space by \(\F\). In DeWitt notation, a condensed index \(A\) represents both
a discrete field label and a spacetime point,
\begin{equation}
A=(a,x),
\qquad
\phi^A=\phi^a(x),
\end{equation}
and repeated indices include both summation and integration:
\begin{equation}
V_AW^A
=
\sum_a\int \dd^dx\,V_a(x)W^a(x).
\end{equation}
A local field redefinition is a map
\begin{equation}
\widetilde\phi^A=\widetilde\phi^A(\phi),
\label{eq:local_redefinition_vdw}
\end{equation}
with invertible functional Jacobian on the domain under consideration. All
statements below are therefore local statements on field-space patches.

The origin of the off-shell non-covariance already appears at the one-loop level, where the Hessian of the classical action enters in the usual trace-log formula. Since the
classical action is a scalar functional on configuration space, its first
functional derivative transforms as a covector:
\begin{equation}
\widetilde S_{,A}
=
\frac{\delta \phi^B}{\delta\widetilde\phi^A}\,
S_{,B}.
\label{eq:first_derivative_transform_vdw}
\end{equation}
The ordinary second derivative, however, is not a tensor:
\begin{equation}
\widetilde S_{,AB}
=
\frac{\delta \phi^C}{\delta\widetilde\phi^A}
\frac{\delta \phi^D}{\delta\widetilde\phi^B}
S_{,CD}
+
\frac{\delta^2\phi^C}{\delta\widetilde\phi^A
\delta\widetilde\phi^B}
S_{,C}.
\label{eq:hessian_nontensor_vdw}
\end{equation}
The second term is the non-tensorial obstruction. It vanishes when the
background satisfies the classical equations of motion, but it is present off-shell. Therefore the ordinary one-loop expression
\begin{equation}
\Gamma_{\rm ord}^{(1)}
=
\frac{i\hbar}{2}\log \Det S_{,AB}
\end{equation}
is not a scalar functional of the background field. This is the usual
parametrization dependence of the one-loop effective action.

The geometrical cure is to replace the ordinary Hessian by the covariant
one.
This construction requires a choice of metric \(G_{AB}\) on configuration
space. In ordinary two-derivative theories, one often reads \(G_{AB}\) from the coefficient of the kinetic term. For instance, if
\begin{equation}
S_{\rm kin}
=
\frac12\int d^dx\,
h_{ij}(\phi)\,\partial_\mu\phi^i\partial^\mu\phi^j ,
\end{equation}
then the corresponding ultralocal field-space metric is
\begin{equation}
G_{(i,x)(j,y)}
=
h_{ij}(\phi(x))\,\delta^{(d)}(x-y).
\end{equation}
In more general theories,
especially higher-derivative theories and EFTs, this prescription need not be
unique. The field-space metric should then be regarded as part of the
defining structures of the quantum theory, together with the action. The present criterion does not select a preferred field-space metric. Instead, \(G_{AB}\) is part of the quantum input rather than being fixed by the action alone. Accordingly, classically related formulations equipped with metrics or measures that are not transported by the same field-space map define different off-shell quantum representations.

Once a metric $G_{AB}$ and the Levi-Civita connection \(\Gamma^C{}_{AB}\) on configuration space are chosen,
one defines
\begin{equation}
S_{;AB}
=
\nabla_A\nabla_B S
=
S_{,AB}
-
\Gamma^C{}_{AB}S_{,C}.
\label{eq:covariant_hessian_vdw}
\end{equation}
The connection term cancels precisely the non-tensorial contribution in
Eq.~\eqref{eq:hessian_nontensor_vdw}. Thus \(S_{;AB}\) is a true covariant
tensor on configuration space, and the trace-log built from the mixed operator
\(G^{AC}S_{;CB}\) has the correct transformation law. In addition to covariant functional derivatives, we stress that the correct covariantization of the Hessian requires the inverse metric factor $G^{AC}$ to raise one configuration-space index. The determinant is indeed only invariant for mixed rank-2 tensors, otherwise transforming as a density for $(0,2)$-type tensors.

\subsection{Covariant source and measure}

The non-covariance of the one-loop effective action is a manifestation of the non-covariant source coupling and measure in Eqs.~\eqref{eq:path_integral_phi_probe} and \eqref{eq:path_integral_linear_source_probe}. The ordinary coupling $J_A \phi^A$ has no invariant meaning because $\phi^A$ is a coordinate, which transforms non-linearly, not a vector. This is more clearly seen in the background field method:
\begin{equation}
    Z[J,\bar\varphi]
    =
    \int\mathcal D \varphi \, \exp i\left\{S[\varphi] + J_A (\varphi^A - \bar\varphi^A)\right\}
    \ ,
    \label{eq:bfm}
\end{equation}
where the current couples to the coordinate difference between the quantum field $\varphi^A$ and the background field $\bar\varphi^A$. Unless the configuration space has a vector structure, the difference $\varphi^A - \bar\varphi^A$ has no intrinsic meaning.

The Vilkovisky--DeWitt construction replaces this coordinate difference by a
geometrical separation. Let
\begin{equation}
\sigma^A(\bar\varphi,\varphi)
\end{equation}
denote the tangent vector at \(\bar\varphi\) to the geodesic that connects
\(\bar\varphi\) to \(\varphi\), with the sign convention chosen so that in a flat
coordinate system
\footnote{
Our sign convention for the geodesic displacement is opposite to the one used in Vilkovisky's
original notation, where in linear coordinates it reads $-\sigma^A(\bar\varphi,\varphi)=\varphi^A-\bar\varphi^A$.
}
\begin{equation}
\sigma^A(\bar\varphi,\varphi)=\varphi^A-\bar\varphi^A .
\end{equation}
Then the source term in Eq.~\eqref{eq:bfm} is written as
\begin{equation}
J_A\,\sigma^A(\bar\varphi,\varphi),
\label{eq:vdw_source_term}
\end{equation}
with \(J_A\) a covector at the base point \(\bar\varphi\). This pairing is now
coordinate invariant. In the scalar example of the previous section, this is
precisely the difference between sourcing \(F(\psi)\), which is the transformed
version of the same observable, and sourcing \(\psi\), which is a different
coordinate-linear observable.

The measure is the other essential ingredient. If configuration space is equipped
with a metric \(G_{AB}(\phi)\), the natural invariant measure is
\begin{equation}
\D\phi\,\sqrt{\Det G_{AB}(\phi)}.
\label{eq:vdw_measure}
\end{equation}
Under a change of field coordinates, the Jacobian from \(\D\phi\) is compensated
by the transformation of \(\sqrt{\Det G}\). In simple perturbative
applications this contribution is often hidden by the choice of coordinates or
by dimensional regularization, but conceptually it is part of the quantum
definition and can affect phenomenology. In gravitational theories, the measure
itself may be organized as part of the effective description, and it can carry
physical information about the configuration-space metric \cite{Falls:2018utl,Kuntz:2025measure}. Related background-independence and field-redefinition covariance issues in quantum gravity are discussed in \cite{Casadio:2022ndh}. The measure is also part of what is required to turn $S_{;AB}$ into the mixed operator $G^{AC}S_{;CB}$ in Eq.~\eqref{eq:covariant_hessian_vdw}.

A covariant source functional may then be written as
\footnote{
Here we are using the background-field generating functional \eqref{eq:bfm}, where the source
is conjugate to the displacement from the base point. For the
ordinary generator with source \(J_A\varphi^A\), the corresponding covariant
source is \(J_A(\bar\varphi^A+\sigma^A)\).
}
\begin{equation}
Z^{\rm VDW}[J,\bar\varphi]
=
\int \D\varphi\,\sqrt{\Det G_{AB}(\varphi)}\,
\exp\left\{
\frac{i}{\hbar}
\left[
S[\varphi]
+
J_A\sigma^A(\bar\varphi,\varphi)
\right]
\right\}.
\label{eq:covariant_generating_functional_vdw}
\end{equation}
We define the connected functional by
\begin{equation}
Z^{\rm VDW}[J,\bar\varphi]
=
\exp\left\{
\frac{i}{\hbar}W[J,\bar\varphi]
\right\},
\label{eq:W_definition_vdw}
\end{equation}
whose functional derivative gives the mean field:
\begin{equation}
\frac{\delta W[J,\bar\varphi]}{\delta J_A}
=
\left\langle
\sigma^A(\bar\varphi,\varphi)
\right\rangle_J
\equiv
\Sigma^A[J,\bar\varphi].
\label{eq:Sigma_definition_vdw}
\end{equation}
Thus the object conjugate to the source is not the coordinate average
\(\langle\varphi^A\rangle\), but the expectation value of the covariant
separation from the base point $\bar\varphi$.

The covariant Legendre transform is then
\begin{equation}
\Gamma[\bar\varphi,\Sigma]
=
W[J,\bar\varphi]
-
J_A\Sigma^A,
\label{eq:covariant_legendre_transform_vdw}
\end{equation}
where \(J_A\) is eliminated in favour of \(\Sigma^A\) through
Eq.~\eqref{eq:Sigma_definition_vdw}. The Vilkovisky--DeWitt effective action is
obtained by choosing the base point $\bar\varphi$ so that the covariant mean field
vanishes:
\begin{equation}
\Sigma^A[J,\bar\varphi]
=
\left\langle
\sigma^A(\bar\varphi,\varphi)
\right\rangle_J
=0.
\label{eq:vdw_mean_field_condition}
\end{equation}
This guarantees that $\bar\varphi=\langle\varphi\rangle$ in linear coordinates.
On this subspace, the Vilkovisky--DeWitt effective action is then defined by
\begin{equation}
\GammaVDW[\bar\varphi]
=
\Gamma[\bar\varphi,0]
=
W[J,\bar\varphi]\big|_{\Sigma=0}.
\label{eq:GammaVDW_from_W}
\end{equation}

Gauge theories require one more step. If \(R_\alpha{}^A(\phi)\) are the
generators of gauge transformations, then the directions tangent to the gauge
orbits are not physical directions in configuration space. The Vilkovisky--DeWitt
construction uses the field-space metric to split tangent vectors into vertical
directions, along the gauge orbits, and horizontal directions, orthogonal to
them. The effective action is then constructed from the horizontal geometry. In
practice one computes with gauge fixing and ghosts, but the final
Vilkovisky--DeWitt effective action is independent of the spurious
gauge-condition dependence that affects the ordinary effective action.

This is particularly important in gravity. The metric field has both a
parametrization ambiguity and a diffeomorphism gauge redundancy. The ordinary
effective action depends on both field coordinates and gauge
condition. The Vilkovisky--DeWitt construction removes these spurious
dependencies by working geometrically on the physical configuration space, or
equivalently by projecting the computation onto horizontal directions in the full
field space.

In the Alice--Bob narrative, should they use the VDW formalism, they would get the same
functional, and hence the same covariant response functions, even without Bob
knowing the field redefinition explicitly. The reason is that the standard prescription of reading the
configuration-space metric from the kinetic term leads Bob to the pulled-back
metric automatically. Indeed, to the accuracy of the EFT, from Eq.~\eqref{eq:free_scalar_redefined_action_probe} Bob would infer
\begin{equation}
G^{(\psi)}_{xy}
=
\left(
1+\frac{6a}{M^2}\psi(x)^2
\right)
\delta^{(d)}(x-y)
+
O(M^{-4}) ,
\end{equation}
which is precisely the pullback of Alice's flat metric
\(G^{(\phi)}_{xy}=\delta^{(d)}(x-y)\). With Bob's metric, the covariant source coupling is also the pullbacks of Alice's. In particular, around \(\bar\psi=0\), the geodesic source
couples to
\begin{equation}
\sigma^\psi(0,\psi)
=
\psi+\frac{a}{M^2}\psi^3
,
\end{equation}
rather than to the coordinate field \(\psi\) itself. The \(O(M^{-2})\) correction
to the ordinary coordinate-linear susceptibility in Eq.~\eqref{eq:chi_psi_vs_chi_phi_direct} is then exactly cancelled by
the nonlinear part of the covariant source coupling. Therefore
\begin{equation}
\chi^{\rm VDW}_\psi(x,y)
=
\chi_\phi(x,y)
+
O(M^{-4}) .
\end{equation}
This equality is not a consequence of the classical actions alone. It relies on
Alice and Bob choosing their field-space metrics consistently with the kinetic term. If the metrics are not defined this way, the comparison would require both metrics to be given along with the actions.

The Vilkovisky--DeWitt effective action will be used as the pivot for the quantum part of our framework. Since it is a scalar functional, and its covariant derivatives are tensors in configuration space, it represents physics in a compact and invariant way. This geometrical effective action does not yet define equivalence, but it supplies the appropriate geometrical objects with which an equivalence criterion can be formulated. To turn the aforementioned tensors into observables, one must still specify what is being measured. This is the role of the probe language introduced next.

\section{Invariant observables}
\label{sec:observables_section}

The basic principle behind our notion of equivalence is operational.
Two descriptions are physically indistinguishable if they yield the same
predictions, and predictions are encoded in observables. A recurrent source of confusion in discussions of field redefinitions is the idea that the fields themselves are observables. They are not. A field variable is part of the theoretical description. What is observable is the outcome of a measurement protocol. In practice one does not measure a bare field component at an abstract point of configuration space. One couples the system to an apparatus, specifies how that coupling is switched on/off and
then reads out a response. Different protocols probe different aspects of the
same dynamics. Hence equivalence must be defined by comparing the scalar quantities produced by
the chosen class of probes.

This is analogous to spacetime physics. The coordinate
component $T_{00}$ is measurable once a frame is fixed, but it has no invariant
meaning. The physical energy density measured by an observer with
four-velocity $u^\mu$ is
\begin{equation}
\rho_u[\Phi]=T_{\mu\nu}[\Phi] \, u^\mu(x) \, u^\nu(x) \, ,
\label{eq:rho_observable_section}
\end{equation}
where $\Phi^A$ includes both the metric and matter fields.
This observable is a scalar obtained by pairing theory data, here $T_{\mu\nu}$, with probe data,
here the observer's four-velocity. Different observers measure different energies, but the
construction itself is invariant. More generally, scalar observables can also take nonlinear forms. For example, the proper time along a prescribed worldline \(\gamma\) is the nonlinear functional
\begin{equation}
\tau_\gamma[g]
=
\int_\gamma d\lambda\,
\sqrt{
-g_{\mu\nu}(z(\lambda))\dot z^\mu(\lambda)\dot z^\nu(\lambda)
}\, .
\label{eq:ptime}
\end{equation}
Here the theory data are represented by the metric \(g_{\mu\nu}\), while the
probe data comprise the worldline \(\gamma\), or equivalently its coordinate
representative \(z^\mu(\lambda)\) on the chosen interval.

Exactly the same logic applies in configuration space. The theory provides
objects such as solutions, correlators and $n$-point functions. The measurement protocol provides the probe data specifying how the apparatus couples to the system. An observable is the scalar functional obtained by
combining the two ingredients, possibly in a nonlinear way. Its value depends on the probe, as it should, but it does not depend on the choice of field coordinates.

We shall denote such a probe-dependent observable by
\begin{equation}
\OO_D[\Phi]\, .
\label{eq:scalarobs}
\end{equation}
Here \(D\) denotes abstractly the data specifying the measurement protocol, \(\OO_D\) is
the corresponding scalar functional on configuration space, and
\(\OO_D[\Phi]\) is its value on the configuration \(\Phi\). Depending on the
problem, \(D\) may include a detector trajectory, a switching function, a
smearing kernel, a hypersurface, asymptotic state data, an initial state, a
source profile, relational clock-and-rod fields, etc.

Under a change of field variables
\(\Phi^A=\Phi^A(\widetilde\Phi)\), the same observable is represented by the
pullback
\begin{equation}
\widetilde\OO_D[\widetilde\Phi]
=
\OO_D[\Phi(\widetilde\Phi)] .
\label{eq:observable_pullback_general}
\end{equation}
We note that \(D\) is the same abstract physical probe in the two descriptions.
If \(D\) has a component representative carrying configuration-space indices,
those components transform tensorially under field redefinitions. By contrast,
spacetime probe data, such as a worldline or hypersurface, are not transformed
by an internal field redefinition, although they do transform under spacetime
diffeomorphisms.

\subsection{Tensorial pairings}

Like Eq.~\eqref{eq:rho_observable_section}, many observables admit a particularly simple representation as tensorial
pairings. In this case the theory supplies a covariant tensor
\(\mathcal T_{A_1\cdots A_n}\) on configuration space, while the probe \(D\)
has a contravariant tensor representative \(D^{A_1\cdots A_n}\). The observable
then takes the form
\begin{equation}
\OO_D[\Phi]
=
\mathcal T_{A_1\cdots A_n}(\Phi)\,
D^{A_1\cdots A_n}\, .
\label{eq:general_pairing_observable_section}
\end{equation}
Here the symbol \(D\) denotes the
abstract physical probe, while \(D^{A_1\cdots A_n}\) denotes its tensorial
representative in the chosen field coordinates.

Equation~\eqref{eq:general_pairing_observable_section} is not the most general
form of an observable. It is the tensorial form appropriate when the protocol
extracts a definite component, or smeared component, of some theory tensor.
Other observables, such as proper time, are scalar functionals that need not be
introduced as finite-rank tensorial pairings, although they may be expanded
locally into such pairings around a chosen background.

The main quantum example is provided by the Vilkovisky--DeWitt effective
action. As reviewed in Section~\ref{sec:vdw_section}, \(\Gamma^{\rm VDW}\) is
a scalar on configuration space. Its covariant derivatives therefore define a
tower of covariant theory tensors,
\begin{equation}
\Gamma^{\rm VDW}_{;A_1\cdots A_n}[\Phi]
=
\nabla_{A_1}\cdots \nabla_{A_n}\GammaVDW[\Phi]\, .
\label{eq:vdw_vertices_observable_section}
\end{equation}
These tensors are the covariant analogues of the usual 1PI vertices. Since they contain all information about the theory, they are
natural objects to pair with tensorial probes as in
Eq.~\eqref{eq:general_pairing_observable_section}.

At the classical level, this tensorial construction becomes relevant when one
defines response observables. The unsourced action \(S[\Phi]\) is already a
scalar functional, and ordinary scalar observables transform by pullback once
the corresponding probe data are specified. However, if one
introduces a source in order to define classical response functions, the usual
coordinate-linear deformation
\begin{equation}
S[\Phi]\longrightarrow S[\Phi]-J_A\Phi^A
\end{equation}
is not intrinsic, for the same reasons outlined in Section~\ref{sec:vdw_section}. A
geometrical approach to the classical response theory parallels the VDW construction, coupling the source to the geodesic
displacement $\sigma(\bar\Phi,\Phi)$.
In this case, the scalar action has the covariant
expansion
\begin{equation}
S[\Phi]
=
S[\bar\Phi]
+
S_{;A}\sigma^A
+
\frac12S_{;AB}\sigma^A\sigma^B
+
\frac1{3!}S_{;ABC}\sigma^A\sigma^B\sigma^C
+\cdots .
\label{eq:covexp}
\end{equation}
Thus classical response coefficients are governed by the covariant derivative
tower \(S_{;A_1\cdots A_n}\). This agrees with the tree-level limit of Eq.~\eqref{eq:vdw_vertices_observable_section}:
\begin{equation}
\Gamma^{\rm VDW}_{;A_1\cdots A_n}
=
S_{;A_1\cdots A_n}
+
O(\hbar).
\end{equation}

\subsection{Gauge-invariant observables}
\label{sec:gauge_dependent_representatives}

In gauge theories and gravity, observables must be gauge invariant \cite{Bergmann:1961wa,Rovelli:1990ph,Dittrich:2005kc}. Thus, if
\(\Phi\) and \(\Phi^g\) are gauge-equivalent configurations, one must have
\begin{equation}
\OO_D[\Phi^g]=\OO_D[\Phi] .
\label{eq:gauge_invariant_observable_general}
\end{equation}
Equivalently, \(\OO_D\) should define a scalar functional on the physical
configuration space \(\F/\Gg\). We note that this condition should be imposed
on the full observable, not on each ingredient appearing in a particular
representative of that observable. In particular, the theory-side quantity
appearing in a formula may be gauge dependent, provided the completed
expression \(\OO_D\) is not.

A simple example illustrates this fact. Let us consider the coupling of an
electromagnetic field \(A_\mu\) to an external current \(J^\mu\) in ordinary
field coordinates,
\begin{equation}
\OO_J[A]
=
\int d^dx\,J^\mu A_\mu .
\label{eq:em_source_probe_example}
\end{equation}
Under \(A_\mu\mapsto A_\mu+\partial_\mu\alpha\), keeping the external current
fixed, one finds
\begin{equation}
\delta_\alpha \OO_J
=
\int d^dx\,J^\mu\partial_\mu\alpha
=
-\int d^dx\,\alpha\,\partial_\mu J^\mu ,
\label{eq:em_source_probe_variation}
\end{equation}
up to boundary terms. Therefore, \(\OO_J[A]\) is gauge invariant only if
\begin{equation}
\partial_\mu J^\mu=0 .
\label{eq:em_current_conservation_probe}
\end{equation}
The current \(J^\mu\) is part of the external protocol. The condition
\eqref{eq:em_current_conservation_probe} says that only conserved current
representatives define this particular gauge-invariant coupling to the gauge
potential.

The same issue appears for response functions. In a gauge-fixed formulation of
Maxwell theory, the photon propagator
\begin{equation}
D_{\mu\nu}^{(\xi)}(k)
=
\frac{-i}{k^2+i\epsilon}
\left[
\eta_{\mu\nu}
-
(1-\xi)\frac{k_\mu k_\nu}{k^2+i\epsilon}
\right]
\label{eq:gauge_fixed_photon_propagator}
\end{equation}
depends on the gauge-fixing parameter \(\xi\). It is therefore not itself a
gauge-invariant observable. However, the scalar response obtained by pairing it
with conserved external currents,
\begin{equation}
\OO_{K,J}
=
K^\mu(k)\,
D_{\mu\nu}^{(\xi)}(k)\,
J^\nu(-k),
\label{eq:conserved_current_response}
\end{equation}
is independent of \(\xi\) whenever
\begin{equation}
k_\mu J^\mu(k)=0,
\qquad
k_\mu K^\mu(k)=0 .
\label{eq:conserved_current_conditions}
\end{equation}
Indeed, the gauge-dependent part of
Eq.~\eqref{eq:gauge_fixed_photon_propagator} is proportional to
\(k_\mu k_\nu\), and its contribution to
Eq.~\eqref{eq:conserved_current_response} is proportional to
\((K\cdot k)(k\cdot J)\), which vanishes. The physical observable is therefore
not the bare gauge-fixed propagator, but the completed scalar response obtained
after pairing it with admissible probe data.

The tensorial pairings introduced above provide the abstract version of this
mechanism. Suppose that an observable is represented as
\begin{equation}
\OO_D[\Phi]
=
\mathcal T_{A_1\cdots A_n}[\Phi]\,
D^{A_1\cdots A_n},
\label{eq:gauge_tensorial_pairing_general}
\end{equation}
where \(\mathcal T_{A_1\cdots A_n}\) is a theory-side tensorial representative
and \(D^{A_1\cdots A_n}\) is the corresponding tensorial representative of the
probe. Under a gauge transformation, the external probe data are kept fixed and
the variation acts on the theory-side representative,
\begin{equation}
\delta_\alpha \OO_D
=
\left(
\delta_\alpha\mathcal T_{A_1\cdots A_n}
\right)
D^{A_1\cdots A_n}.
\label{eq:gauge_variation_tensorial_pairing_general}
\end{equation}
Thus the gauge-invariance condition is
\begin{equation}
\left(
\delta_\alpha\mathcal T_{A_1\cdots A_n}
\right)
D^{A_1\cdots A_n}
=
0
\label{eq:gauge_invariance_tensorial_pairing_general}
\end{equation}
for arbitrary gauge parameters. This condition is imposed on the contraction as
a whole. The tensor \(\mathcal T_{A_1\cdots A_n}\) need not be gauge invariant
by itself; what matters is whether the pairing defines a scalar functional on
\(\F/\Gg\).

A useful way of implementing this condition is to choose probe representatives
that annihilate vertical directions. Let \(R_\alpha{}^A\) denote the generators
of gauge transformations on field space and, when a field-space metric
\(G_{AB}\) has been chosen, define
\begin{equation}
R_{\alpha A}=G_{AB}R_\alpha{}^B .
\end{equation}
Then the vertical-annihilation condition is
\begin{equation}
R_{\alpha A_k}\,
D^{A_1\cdots A_k\cdots A_n}
=
0
\qquad
\text{for each slot } k .
\label{eq:horizontal_probe_condition}
\end{equation}
This condition should be understood as an admissibility condition on the
tensorial representative of the probe, relative to a gauge-redundant
representative of the theory data. It is the field-space analogue of current
conservation in Eq.~\eqref{eq:em_current_conservation_probe}.

The restrictions on the probes should not be read as restrictions on which
experiments can be performed. Rather, they specify which protocols test a given
invariant sector of the theory through a chosen representative. A
gauge-dependent theory tensor may still be useful, but it becomes part of an
observable only after the probe data, projection, dressing, boundary conditions
or relational structure needed for gauge invariance have been included.

Thus the relevant distinction is between gauge-dependent representatives and
gauge-invariant observables. Field variables, propagators and response tensors
may depend on a gauge choice. The observable \(\OO_D\), obtained after the full
protocol has been specified, must not. Restrictions on the representative of
\(D\) are useful because they serve as admissibility conditions for using
gauge-redundant theory data to construct a scalar observable on \(\F/\Gg\).

\subsection{Probe spaces}
\label{sec:probe}

We now distinguish different classes of probes. This distinction is needed
because different classes test different amounts of information about a theory.
Some probe classes are mathematical devices containing all possible mathematically well-defined probes, which are used to infer mathematical equivalence. Others are physical classes, adapted to a particular type of
measurement, such as scattering and detector responses.

Two independent distinctions will be useful. First, some probes are tensorial as
they contract with theory tensors to form scalars. Others define scalar observables, but do
not arise naturally as tensor pairings. Second, a probe class can be
separating, in the mathematical sense that it distinguishes the theory tensors
under consideration, or physically motivated, in the sense that it represents the
measurements available in a given setup. A physically natural class need not be
separating, and a separating class need not be physically realizable.

We first consider a mathematical idealization. Suppose the object to be tested
is a theory tensor \(\mathcal T_{A_1\cdots A_n}\). One may then consider
all tensorial probes \(D^{A_1\cdots A_n}\) for which the pairing
\begin{equation}
\OO_D[\Phi]
=
\mathcal T_{A_1\cdots A_n}(\Phi)
D^{A_1\cdots A_n}
\label{eq:test_probe_pairing}
\end{equation}
is well defined. We denote this large class by
\begin{equation}
\mathfrak D_{\rm test}.
\end{equation}
The rank of \(D^{A_1\cdots A_n}\) is understood from the tensor being tested.

The purpose of \(\mathfrak D_{\rm test}\) is mathematical separation. If two
theory tensors \(\mathcal T^{(1)}\) and \(\mathcal T^{(2)}\) give the same
scalar for every test probe,
\begin{equation}
\mathcal T^{(1)}_{\ A_1\cdots A_n}D^{A_1\cdots A_n}
=
\mathcal T^{(2)}_{\ A_1\cdots A_n}D^{A_1\cdots A_n}
\qquad
\text{for every }D\in\mathfrak D_{\rm test},
\label{eq:test_probe_separation_condition}
\end{equation}
then
\begin{equation}
\mathcal T^{(1)}_{\ A_1\cdots A_n}
=
\mathcal T^{(2)}_{\ A_1\cdots A_n}
\end{equation}
as tensorial distributions. In ordinary QFT, this is the familiar statement
that distributions are determined by their action on test functions.
Mathematical separation therefore says that the full collection of pairings
with test probes determines the tensor \(\mathcal T\) uniquely.

The test class should not be understood as the universal class of all possible
probes. It is tied to tensorial pairings such as
Eq.~\eqref{eq:test_probe_pairing}. Some physical observables, such as proper time, are scalar functionals rather
than contractions with a fixed-rank theory tensor.
They are therefore not naturally elements of \(\mathfrak D_{\rm test}\),
although they may be expanded locally into tensorial coefficients around a
chosen background.

There is a corresponding separation notion for scalar observables of this kind. A family
\(\mathfrak D_{\rm sep}\) of probes entering scalar functionals \(\OO_D\) may
be called separating if
\begin{equation}
\OO_D[\Phi_1]=\OO_D[\Phi_2]
\qquad
\text{for every }D\in\mathfrak D_{\rm sep}
\label{eq:scalar_probe_separation_condition}
\end{equation}
implies
\begin{equation}
\Phi_1=\Phi_2 .
\end{equation}
In the presence of gauge redundancies, the corresponding condition is
separation of gauge orbits,
\begin{equation}
\OO_D[\Phi_1]=\OO_D[\Phi_2]
\qquad
\text{for every }D\in\mathfrak D_{\rm sep}
\quad
\Longrightarrow
\quad
[\Phi_1]=[\Phi_2]\in \F/\Gg .
\label{eq:scalar_probe_gauge_orbit_separation}
\end{equation}
Thus \(\mathfrak D_{\rm sep}\) is the nonlinear analogue of
\(\mathfrak D_{\rm test}\). The former separates configurations, or gauge
orbits, while the latter separates theory tensors.

In practice, the relevant probe class is fixed by the measurement being modeled. It collects the probes
that can actually be prepared, controlled or read out in that setup. Such classes are usually smaller
than the mathematical separating classes. Indeed, real detectors have finite size,
finite switching time and limited resolution. Scattering experiments prepare
wave packets rather than exact momentum eigenstates. Cosmological observables
are tied to specific clocks, rods and coarse-graining procedures. We therefore write, depending on the type of
observable,
\begin{equation}
\mathfrak D_{\rm phys}\subset \mathfrak D_{\rm test}
\qquad \text{or} \qquad
\mathfrak D_{\rm phys}\subset \mathfrak D_{\rm sep}.
\end{equation}
When gauge redundancies are present, physical admissibility also includes the
requirement that observables \(\OO_D\) be gauge invariant, as discussed in
Section~\ref{sec:gauge_dependent_representatives}.
We also note that physical classes need not be separating. Such a class is separating only if it is large enough to
distinguish the relevant configurations, gauge orbits or tensorial
distributions.

We now spell out the main physical probe classes used in the rest of the paper.
The point of the following examples is not to give an exhaustive taxonomy, but
to make explicit what the element \(D\) contains in each case. The simplest one is the class of classical kinematical probes,
\begin{equation}
\mathfrak D_{\rm kin}^{\rm cl}
\subset
\mathfrak D_{\rm sep}.
\label{eq:clkin}
\end{equation}
An element \(D\in\mathfrak D_{\rm kin}^{\rm cl}\) specifies the kinematical data needed to
define a scalar observable
\begin{equation}
\OO_D:\F\longrightarrow \mathbb R .
\end{equation}
For example, the proper-time observable \eqref{eq:ptime} is specified by data such as
\begin{equation}
D_\tau=(\gamma,I),
\end{equation}
where \(\gamma:I\to M\) is the worldline of the clock. Similarly, a
charge-type observable may be specified by hypersurface and smearing data,
\begin{equation}
D_Q=(\Sigma,\eta),
\end{equation}
with
\begin{equation}
\OO_{D_Q}[\Phi]
=
\int_\Sigma \eta\, j^\mu[\Phi]\,d\Sigma_\mu .
\label{eq:charge_probe_readout}
\end{equation}
The detailed content of \(D\) thus depends on the protocol. It may include a
worldline, a hypersurface, a smearing kernel, a detector trajectory or
relational clock-and-rod data. The
field-space metric may be part of the representation of the theory, but it is
not tested by this class of probes.

A second important choice is the class of classical response probes,
\begin{equation}
\mathfrak D_{\rm resp}^{\rm cl}
\subset
\mathfrak D_{\rm test}.
\label{eq:clresp}
\end{equation}
Here the probe is tensorial. For an \(n\)-th order response observable, an
element of the class has a representative
\begin{equation}
D=D^{A_1\cdots A_n},
\end{equation}
and the readout is the scalar pairing
\begin{equation}
\OO_D^{(n)}[\bar\Phi]
=
\mathcal R_{A_1\cdots A_n}[\bar\Phi]\,
D^{A_1\cdots A_n}.
\label{eq:classical_response_probe_pairing}
\end{equation}
The response tensor \(\mathcal R_{A_1\cdots A_n}\) is defined by introducing a
covariant source,
\begin{equation}
S_J[\Phi;\bar\Phi]
=
S[\Phi]-J_A\sigma^A(\bar\Phi,\Phi),
\label{eq:classical_response_probe_source}
\end{equation}
solving the sourced equations of motion, and differentiating the sourced
solution, or a scalar readout evaluated on that solution, with respect to
\(J_A\). Schematically, for a chosen scalar readout \(Q[\Phi]\),
\begin{equation}
\mathcal R^Q_{A_1\cdots A_n}[\bar\Phi]
=
\left.
\frac{\delta^n Q[\Phi_J]}
{\delta J_{A_1}\cdots \delta J_{A_n}}
\right|_{J=0}.
\label{eq:classical_response_tensor}
\end{equation}
The explicit response coefficients are determined by the covariant derivative
tower \(S_{;A_1\cdots A_k}\). The tensor
\(D^{A_1\cdots A_n}\) encodes the physically allowed smearing, switching,
support and readout profile. Since this construction uses
\(\sigma^A(\bar\Phi,\Phi)\), response probes test the field-space geometry
\(G_{AB}\), not only the unsourced action.

At the quantum level, a particularly important physical class is the scattering
class,
\begin{equation}
\mathfrak D_{\rm scatt}.
\end{equation}
An element of this class is a pair of asymptotic data,
\begin{equation}
D=(D_{\rm in},D_{\rm out}),
\end{equation}
where, schematically,
\begin{equation}
D_{\rm in}
=
\left\{(f_i,\alpha_i)\right\}_{i=1}^{m},
\qquad
D_{\rm out}
=
\left\{(h_j,\beta_j)\right\}_{j=1}^{n}.
\end{equation}
The functions \(f_i\) and \(h_j\) are incoming and outgoing wave packets, while
\(\alpha_i\) and \(\beta_j\) denote all quantum numbers (e.g. species, spin, polarization etc). The associated observable is the transition
amplitude
\begin{equation}
\OO_D
=
\langle D_{\rm out};{\rm out}|D_{\rm in};{\rm in}\rangle .
\label{eq:scattering_probe_observable}
\end{equation}
The standard equivalence theorem may be viewed as equivalence relative to
\(\mathfrak D_{\rm scatt}\). In LSZ language, the scattering probe packages the external-state data together
with the amputation and on-shell projection. The theorem says that local invertible field
redefinitions leave the scalar obtained after applying this probe unchanged,
even though the off-shell Green functions themselves need not agree. Thus
\(\mathfrak D_{\rm scatt}\) tests only the on-shell asymptotic sector of the
theory, not its full off-shell tensorial content.

Scattering, however, is only one quantum probe class. If the aim is to compare
dynamical or off-shell information, one must allow a larger class,
\begin{equation}
\mathfrak D_{\rm off\text{-}shell}.
\end{equation}
One type of element is a tensorial probe \(D^{A_1\cdots A_n}\) paired with
covariant effective vertices,
\begin{equation}
\OO_D^{(n)}[\Phi]
=
(\GammaVDW)_{;A_1\cdots A_n}[\Phi]\,
D^{A_1\cdots A_n}.
\label{eq:offshell_vdw_probe_pairing}
\end{equation}
Another type is a state-dependent correlator probe,
\begin{equation}
D=(\rho,\mathcal C,K^{AB}),
\end{equation}
where \(\rho\) is the state or density matrix, \(\mathcal C\) denotes the
choice of contour or Green-function prescription, and \(K^{AB}\) is the
smearing kernel. The corresponding readout has the schematic form
\begin{equation}
\OO_D[\Phi]
=
K^{AB}G^{(\rho,\mathcal C)}_{AB}[\Phi],
\label{eq:offshell_state_dependent_probe_pairing}
\end{equation}
where \(G^{(\rho,\mathcal C)}_{AB}\) may be Wightman, retarded, Euclidean or
in-in, depending on the physical setup. Such probes test off-shell information
and are therefore stronger than scattering probes. Equality of the off-shell
VDW tensorial content implies equality of on-shell amplitudes only after the
asymptotic structures and the LSZ projection have also been matched.

The distinction between \(\mathfrak D_{\rm test}\), \(\mathfrak D_{\rm sep}\)
and \(\mathfrak D_{\rm phys}\) is not a single chain of inclusions. These
classes answer different questions. The class \(\mathfrak D_{\rm test}\)
separates theory tensors. The class \(\mathfrak D_{\rm sep}\) separates
configurations or gauge orbits by scalar readouts. The class
\(\mathfrak D_{\rm phys}\) specifies which probes are physically admitted in
the comparison. Depending on the problem, \(\mathfrak D_{\rm phys}\) may be
\(\mathfrak D_{\rm kin}^{\rm cl}\), \(\mathfrak D_{\rm resp}^{\rm cl}\),
\(\mathfrak D_{\rm scatt}\), a more general on-shell quantum class, or an
off-shell class \(\mathfrak D_{\rm off\text{-}shell}\).

This distinction also controls the amount of geometrical structure tested by a
classical comparison. For non-response readouts, such as proper time or an
integrated charge, the observable is already a scalar functional once the
physical probe is specified. The field-space metric is therefore not tested by
this class of readouts. For response probes, however, the metric and connection
enter the definition of the covariant source and of the response tensors.
Therefore, if two classical descriptions are to be equivalent relative to a
response-probe class, their field-space geometries cannot be chosen
independently. Otherwise the two descriptions define different response
experiments, even if their unsourced classical actions are related by a field
redefinition.

Phrasing the problem in terms of probe classes allows for
different levels of equivalence. 
If equality is imposed only for a restricted
physical class, such as \(\mathfrak D_{\rm scatt}\), two descriptions may be
equivalent for the corresponding experiments while differing off shell. If
equality is imposed for a separating test class, then the corresponding
tensorial or configurational data agree. The probe class therefore controls
both what is being compared and how strong the resulting notion of equivalence
is.

\subsection{A detector example}

Consider an Unruh--DeWitt detector \cite{Unruh:1976db,DeWitt:1979ue}, a two-level
system with energy gap $\Omega$ coupled linearly to a quantum scalar field along a worldline $z(\tau)$, with switching
function $\chi(\tau)$. To leading nontrivial order in the coupling, the excitation probability is
controlled by the response function
\begin{equation}
F(\Omega)
=
\int \dd\tau\,\dd\tau'\,
\chi(\tau)\chi(\tau')\,e^{-i\Omega(\tau-\tau')}
W\!\bigl(z(\tau),z(\tau')\bigr)\, ,
\label{eq:udw_response_observable_section}
\end{equation}
where $W(x,x')$ is the Wightman function in the chosen quantum state.

The response \eqref{eq:udw_response_observable_section} is a special case of the general pairing
\eqref{eq:general_pairing_observable_section} with $n=2$. Defining the distributional probe
\begin{equation}
F_\Omega(x)
=
\int \dd\tau\,
\chi(\tau)e^{i\Omega\tau}
\delta^{(d)}\!\bigl(x-z(\tau)\bigr)\, \ ,
\label{eq:Fomega_observable_section}
\end{equation}
Eq.~\eqref{eq:udw_response_observable_section} can be written as
\begin{equation}
F(\Omega)
=
\int \dd^dx\,\dd^dx'\,
\overline{F_\Omega(x)}\,
W(x,x')\,
F_\Omega(x')\, .
\label{eq:udw_pairing_observable_section}
\end{equation}
The response is a probability, hence a modulus square of an amplitude, so it
naturally contains the probe and its complex conjugate.

This example is valuable for the present paper for two reasons. First, it is an observable that is
manifestly not an on-shell scattering amplitude. Second, it is explicitly built from a measurement
protocol. The probe contains the detector trajectory \(z(\tau)\), the switching function \(\chi(\tau)\),
and the detector energy gap \(\Omega\). The observable is not the bare Wightman function by itself,
but its pairing with a physical apparatus. In
gravity and in curved-space QFT this is often the natural kind of observable one actually wants to
compare across different formulations.

The same language also accommodates relational observables in gravity\footnote{We do not attempt to construct a complete algebra of gravitational
observables. We assume only that an admissible class of
diffeomorphism-invariant or relational probes has been specified.} \cite{Rovelli:1990ph,Dittrich:2005kc,Westman:2007events,Westman:2007coordinates}. A local
expression \(f(x)\) is not a diffeomorphism-invariant
observable, because the spacetime label \(x\) has no invariant meaning. One may
instead introduce scalar reference fields \(X^\mu(x)\), interpreted as physical
rods and clocks, and restrict to a region where they define a good local
coordinate system,
\begin{equation}
\det\!\left(\frac{\partial X^\mu}{\partial x^\nu}\right)\neq0 .
\end{equation}

For a scalar quantity \(f(x)\), a relational observable may then be written as
\begin{equation}
\OO_f(\sigma)
=
\int d^4x\,
f(x)\,
\delta^{(4)}(X(x)-\sigma)
\left|
\det\left(\frac{\partial X}{\partial x}\right)
\right| .
\label{eq:relational_observable_section}
\end{equation}
Indeed, if \(x_\sigma\) is defined by
\(X^\alpha(x_\sigma)=\sigma^\alpha\), then
\begin{equation}
\OO_f(\sigma)=f(x_\sigma).
\end{equation}
Equivalently, using
\begin{equation}
\det\!\left(
g^{\mu\nu}\partial_\mu X^\alpha\partial_\nu X^\beta
\right)
=
\frac{
\left[\det(\partial X/\partial x)\right]^2
}{\det g},
\end{equation}
one may write the same observable as
\begin{equation}
\OO_f(\sigma)
=
\int d^4x\sqrt{-g}\,
f(x)\,
D_\sigma[g,X](x),
\label{eq:relational_pairing}
\end{equation}
where
\begin{equation}
D_\sigma[g,X](x)
=
\delta^{(4)}(X(x)-\sigma)
\sqrt{
\left|
\det\left(
g^{\mu\nu}\partial_\mu X^\alpha\partial_\nu X^\beta
\right)
\right|
}.
\label{eq:relational_probe_density}
\end{equation}
Thus \(D_\sigma[g,X]\) is the scalar localization factor that converts the
invariant volume element \(d^4x\sqrt{-g}\) into the relational point evaluation.
Eq.~\eqref{eq:relational_pairing} is again a probe pairing, but now the probe is dynamical rather than
external because it is built from the dynamical fields \(X^\mu\). We do not need relational probes in every
application since the external ones are perfectly adequate in many cases.
Nonetheless, Eq.~\eqref{eq:relational_observable_section} shows that the
pairing language is broad enough to include both types of situations.
This is important in gravity, where the use of
non-dynamical background structures is often undesirable.

\section{Classical equivalence}
\label{sec:classical_equiv_section}

We now turn the previous examples and the observable language into a classical
criterion of equivalence. The main point is that the classical data to be
compared depend on the chosen probe class. For finite scalar readouts, the
minimal data $(\F,\Gg,S)$ are sufficient. If the probe class includes response
functions, the description must also include the field-space geometry used to
define covariant sources and the tensors $S_{;A_1\cdots A_n}$. With this
distinction in mind, we first define classical descriptions, then direct local
equivalence, parent-mediated equivalence, and finally the distinction between
local, branchwise and global equivalence.

\subsection{Direct local classical equivalence}
\label{sec:direct_local_classical_equivalence}

A classical representation will be denoted by
\begin{equation}
\MM_{\rm cl}
=
(\F,\Gg,S,G)\, .
\label{eq:classical_representation}
\end{equation}
Here \(\F\) is the configuration space, \(\Gg\) is the gauge group when the
theory has gauge redundancies, \(S\) is the classical action and \(G_{AB}\) is a
metric on configuration space. The inclusion of \(G_{AB}\) in
\(\MM_{\rm cl}\) does not mean that every classical probe class is sensitive to
it. Rather, the probe class determines which observables are being compared and, through their construction, which
structures of the representation must be transported.

Let
\begin{equation}
\MM_{{\rm cl},1}
=
(\F_1,\Gg_1,S_1,G_1),
\qquad
\MM_{{\rm cl},2}
=
(\F_2,\Gg_2,S_2,G_2)
\end{equation}
be two classical representations. We say that they are directly locally equivalent relative to a probe class
\(\mathfrak D\) on a domain \(U_2\subset\F_2\) if there exists a local
diffeomorphism
\begin{equation}
    \Phi:U_2\subset\F_2\longrightarrow U_1\subset\F_1
\end{equation}
such that, for every probe \(D\in\mathfrak D\), all corresponding scalar
observables agree,
\begin{equation}
    \OO^{(2)}_D[\varphi]
    =
    \OO^{(1)}_D[\Phi(\varphi)]
    \qquad
    \text{for all }\varphi\in U_2 .
    \label{eq:scalarequiv}
\end{equation}
Here the same symbol \(D\) denotes the same physical probe, which however is represented differently in different field coordinates.

This operational definition is not usually the most practical criterion, since
it would require comparing all observables in the class. In practice, it is
enough to require that all ingredients entering the construction of
\(\OO_D\) be transported by \(\Phi\). These ingredients include the probe data
\(D\) themselves, as well as the theory-side structures needed to construct the
object probed by \(D\). There are very few such structures in \eqref{eq:classical_representation}, hence the number of useful probe classes reduces drastically, only including $\mathfrak D_{\rm kin}^{\rm cl}$ (Eq.~\eqref{eq:clkin}) and $\mathfrak D_{\rm resp}^{\rm cl}$ (Eq.~\eqref{eq:clresp}).

For the classical probe classes considered here, the basic theory-side
ingredient is the dynamics. A sufficient structural condition for transporting
the dynamics is the pullback of the action,
\begin{equation}
S_2[\varphi]
=
S_1[\Phi(\varphi)]
\qquad
\text{for all } \varphi\in U_2\, .
\label{eq:direct_local_action_equivalence}
\end{equation}
In an EFT, Eq.~\eqref{eq:direct_local_action_equivalence} is understood only up
to the stated truncation order. If gauge redundancies are present, the map must
also preserve gauge orbits. Equivalently, there should exist a group
isomorphism \(\psi:\Gg_2\to\Gg_1\) such that
\begin{equation}
\Phi(g\cdot\varphi)
=
\psi(g)\cdot\Phi(\varphi)\, .
\label{eq:gauge_equivariance_classical_section}
\end{equation}
This ensures that configurations identified in one description are mapped to
configurations identified in the other.

From Eq.~\eqref{eq:direct_local_action_equivalence}, it follows that
\begin{equation}
\frac{\delta S_2}{\delta\varphi^A}
=
\frac{\delta\Phi^B}{\delta\varphi^A}
\frac{\delta S_1}{\delta\Phi^B}\, ,
\label{eq:eom_covariance_classical_direct}
\end{equation}
and hence solutions are mapped into solutions whenever the Jacobian of
\(\Phi\) is invertible. Moreover, any scalar observable
\(\OO_D^{(1)}\) in the first description has a transported representative in
the second description,
\begin{equation}
\OO_D^{(2)}[\varphi]
=
\OO_D^{(1)}[\Phi(\varphi)]\, .
\label{eq:readout_transport_classical_direct}
\end{equation}
Thus, for the probe class $\mathfrak D_{\rm kin}^{\rm cl}$, whose elements define kinematical readouts rather than response functions, no further theory-side structure
is needed. The probe data \(D\) specify the readout, while the action determines
the dynamics. Therefore Eq.~\eqref{eq:direct_local_action_equivalence}, together
with the transport of the probe data themselves, is sufficient. The field-space
metric \(G_{AB}\) is not tested by this class.

For $\mathfrak D_{\rm resp}^{\rm cl}$ (defined in Eq.~\eqref{eq:clresp}), which includes response probes, the equivalence criterion is
stronger. In that case the field-space metric also enters the observable
construction, because it fixes the geodesic displacement
\(\sigma^A(\bar\Phi,\Phi)\) and the covariant derivative tower of the action.
One must therefore require
\begin{equation}
G_2
=
\Phi^\ast G_1\, .
\label{eq:geometry_pullback_classical_direct}
\end{equation}
In components,
\begin{equation}
(G_2)_{AB}(\varphi)
=
\frac{\delta\Phi^C}{\delta\varphi^A}
\frac{\delta\Phi^D}{\delta\varphi^B}
(G_1)_{CD}(\Phi(\varphi))\, .
\label{eq:metric_pullback_classical_direct}
\end{equation}
Since the connection is assumed to be Levi-Civita, Eq.~\eqref{eq:geometry_pullback_classical_direct}
also transports the connection.

With this additional condition, the covariant derivative towers of the two
actions are related tensorially:
\begin{equation}
(S_2)_{;A_1\cdots A_n}(\varphi)
=
\frac{\delta\Phi^{B_1}}{\delta\varphi^{A_1}}
\cdots
\frac{\delta\Phi^{B_n}}{\delta\varphi^{A_n}}
(S_1)_{;B_1\cdots B_n}(\Phi(\varphi))\, .
\label{eq:covariant_action_tower_classical_direct}
\end{equation}
Consequently, covariant source deformations and the corresponding response
functions are transported consistently between the two descriptions. If the
actions are related by Eq.~\eqref{eq:direct_local_action_equivalence} but the
metrics are not related by Eq.~\eqref{eq:geometry_pullback_classical_direct},
the two representations may still be equivalent for finite scalar readouts, but
they are not equivalent relative to a response-probe class. They then define
different covariant source experiments.

\subsection{Parent-mediated classical equivalence}
\label{sec:parent_mediated_classical_equivalence}

Direct local equivalence is appropriate when the field content is unchanged. It
does not, however, describe all standard reformulations of a theory.
The metric \(f(R)\) example of Sec.~\ref{sec:fr_seed} is the prototype. The
metric endpoint has configuration space \(\F_g\). The scalar--tensor
description is obtained only after passing through an enlarged configuration
space, for instance
\begin{equation}
\F_{\rm ext}
=
\F_g\times\F_\chi\times\F_\lambda
\end{equation}
in the Lagrange-multiplier formulation, or
\begin{equation}
\widehat\F
=
\F_g\times\F_\chi
\end{equation}
after eliminating \(\lambda\). The recovery of the metric \(f(R)\) endpoint is
not a change of coordinates on \(\F_g\). One rather restricts the variables in $\mathcal F_{\rm ext}$ (or $\widehat\F$) to the original variables in $\F_g$ via the equations of motion. By contrast, the passage
from the Jordan variables \((g,\phi)\) to the Einstein variables
\((\hat g,\sigma)\) is a genuine local reparametrization once the scalar--tensor
field space has already been introduced. Parent-mediated equivalence is the
abstract version of the former notion.

A parent construction begins by enlarging the configuration space. Starting
from an endpoint representation
\begin{equation}
\MM_{\rm cl}^{\rm end}
=
(\F_{\rm end},\Gg_{\rm end},S_{\rm end},G_{\rm end}),
\end{equation}
one introduces additional variables and considers a parent representation
\begin{equation}
\widetilde{\MM}_{\rm cl}
=
(\widetilde\F,\widetilde\Gg,\widetilde S,\widetilde G),
\qquad
\widetilde\F
=
\F_{\rm end}\times\F_{\rm aux},
\label{eq:parent_extension_classical}
\end{equation}
together with the canonical projection
\begin{equation}
\pi:\widetilde\F\longrightarrow\F_{\rm end}.
\label{eq:parent_projection_classical}
\end{equation}
The extra variables may be introduced for different reasons. In an auxiliary
extension they are used to lower derivative order, linearize an interaction, or
rewrite the dynamics in a more convenient form. In a gauge-completing extension,
such as the St\"uckelberg construction, the enlarged field space carries a larger
gauge redundancy and the endpoint is recovered by gauge fixing or by passing to
a quotient. In either case, the endpoint is recovered by reduction on the
enlarged space, not by a diffeomorphism between the original endpoint field
spaces.

The word ``parent'' is therefore relational. A representation is a parent only
relative to specified endpoint data and specified reduction maps. Likewise, an
added variable is not intrinsically auxiliary. It is auxiliary only relative to
a chosen endpoint. This is exactly what happens in the \(f(R)\) chain, where the
scalar introduced in the parent is eliminated relative to the metric endpoint,
but it becomes part of the field content in the scalar--tensor endpoint.

Let the parent variables be written as
\begin{equation}
\widetilde\Phi=(\Phi_{\rm end},\Xi),
\end{equation}
where \(\Phi_{\rm end}\) are the variables of the endpoint under consideration
and \(\Xi\) denotes the auxiliary variables to be eliminated relative to that endpoint. If,
on a domain \(U\subset\F_{\rm end}\), the equations
\begin{equation}
\left.
\frac{\delta\widetilde S}{\delta\Xi}
\right|_{\Xi=\Xi_\star(\Phi_{\rm end})}
=0
\label{eq:aux_eom_parent_classical}
\end{equation}
are locally solvable and nondegenerate, they define a section
\begin{equation}
s:U\longrightarrow\widetilde\F,
\qquad
\pi\circ s=\mathrm{id}_U ,
\label{eq:parent_section_classical}
\end{equation}
given by
\begin{equation}
    s(\Phi_{\rm end})
    =
    (\Phi_{\rm end},\Xi_\star(\Phi_{\rm end})).
\end{equation}
The reduced endpoint action is the pullback of the parent action,
\begin{equation}
S_{\rm end}
=
s^\ast\widetilde S
\qquad
\text{on } U ,
\label{eq:parent_reduced_action_classical}
\end{equation}
or more explicitly:
\begin{equation}
    S_{\rm end}[\Phi_{\rm end}]
    =
    \widetilde S[\Phi_{\rm end},\Xi_\star(\Phi_{\rm end})]
    \qquad
    \text{on } U .
\end{equation}
For the metric \(f(R)\) endpoint, this is the abstract form of the statement
\begin{equation}
S_f[g]
=
\widehat S[g,\chi=R[g]]
=
s_f^\ast \widehat S[g,\chi].
\end{equation}
In a gauge-completing extension, the section \(s\) may represent a local gauge choice, or the reduction
may be formulated invariantly by passing to the corresponding quotient. In all
cases, the domain \(U\) and the chosen reduction branch are part of the equivalence data. Changing either one changes the local equivalence statement.

Endpoint descriptions may also be obtained after reparametrizing the parent
fields. Thus a parent construction may contain maps
\begin{equation}
\mathscr P:W\subset\widetilde\F\longrightarrow W'\subset\widetilde\F
\label{eq:parent_space_reparametrization}
\end{equation}
on the enlarged configuration space, followed by reductions to different
endpoints. This is why the parent relation is naturally represented by a local
zig--zag,
\begin{equation}
\MM_{{\rm cl},1}
\;\longleftarrow\;
\widetilde{\MM}_{\rm cl}
\;\longrightarrow\;
\MM_{{\rm cl},2},
\label{eq:classical_parent_zigzag}
\end{equation}
where each leg is a reduction, a gauge reduction, a parent-space
reparametrization, or a composition of such operations. The arrow notation does
not mean that the endpoint field spaces are directly diffeomorphic. It means
that both endpoint descriptions are obtained from a common enlarged
representation by specified local operations.

We can now state the parent-mediated criterion in operational form. Two
endpoint representations
\begin{equation}
\MM_{{\rm cl},1}
=
(\F_1,\Gg_1,S_1,G_1),
\qquad
\MM_{{\rm cl},2}
=
(\F_2,\Gg_2,S_2,G_2)
\end{equation}
are parent-mediated classically equivalent relative to a probe class
\(\mathfrak D\) on domains \(U_i\subset\F_i\) if there exists a parent
representation \(\widetilde{\MM}_{\rm cl}\) and local endpoint legs
\begin{equation}
s_i:U_i\subset\F_i\longrightarrow\widetilde\F,
\qquad
i=1,2,
\end{equation}
possibly composed with parent-space reparametrizations, such that, for every
probe \(D\in\mathfrak D\), there exists a parent scalar observable
\(\widetilde\OO_D\) whose endpoint representatives are
\begin{equation}
\OO_D^{(i)}
=
s_i^\ast\widetilde\OO_D,
\qquad
i=1,2 .
\label{eq:parent_operational_equivalence_classical}
\end{equation}
Equivalently,
\begin{equation}
\OO_D^{(i)}[\varphi_i]
=
\widetilde\OO_D[s_i(\varphi_i)],
\qquad
\varphi_i\in U_i .
\label{eq:parent_observable_pullback_classical}
\end{equation}
Equality of endpoint observables
is therefore mediated by the parent. Endpoint configurations are compared when
their representatives in \(\widetilde\F\) describe the same parent
configuration.

As in Section~\ref{sec:direct_local_classical_equivalence}, the operational
definition can be implemented by transporting the ingredients that enter
the construction of the observables. For the kinematical probe class
\(\mathfrak D_{\rm kin}^{\rm cl}\), the relevant theory-side ingredient is the
dynamics. A sufficient structural condition is therefore
\begin{equation}
S_i=s_i^\ast\widetilde S,
\qquad
i=1,2 ,
\label{eq:parent_action_equivalence_classical}
\end{equation}
where $S_i$ denotes the endpoint actions and $\widetilde S$ the parent one.
For the response probe class \(\mathfrak D_{\rm resp}^{\rm cl}\), the parent-mediated analogue
of Eq.~\eqref{eq:geometry_pullback_classical_direct} is
\begin{equation}
G_i=s_i^\ast\widetilde G,
\qquad
i=1,2,
\label{eq:parent_response_equivalence_classical}
\end{equation}
for endpoint metrics $G_i$ and parent $\widetilde G$. If
Eq.~\eqref{eq:parent_response_equivalence_classical} is not specified, the
parent construction defines equivalence only on \(\mathfrak D_{\rm kin}^{\rm cl}\), not a response-level equivalence on \(\mathfrak D_{\rm resp}^{\rm cl}\).

Our formalism thus separates two questions that are often
conflated. One question is whether two endpoint descriptions are obtained from a
common parent on the domains of interest. Another is whether there exists a
direct field redefinition between the endpoint variables themselves. The first
may hold even when the second does not. Metric \(f(R)\) gravity and its
scalar--tensor representation provide the basic example of parent-mediated equivalence. The Jordan--Einstein step, on the other hand, is a direct local field
redefinition.

\subsection{Local, branchwise and global classical equivalence}
\label{sec:local_branchwise_global_classical_equivalence}

The equivalence criteria above are local by construction. A direct map
\(\Phi:U_2\to U_1\), or a parent reduction section \(s:U\to\widetilde\F\),
need only exist on a domain of configuration space. This is indeed what happened in the examples discussed in
Sec.~\ref{sec:fr_seed} since the relation \(\phi=f'(\chi)\) requires
\(f''(\chi)\neq0\), and the Einstein-frame variables require \(\phi>0\).
The domain on which the map is defined is therefore part of
the equivalence statement.

Let us first consider a direct relation. Suppose that a domain
\(U_2\subset\F_2\) is covered by patches
\begin{equation}
U_2=\bigcup_\alpha U_{2,\alpha},
\end{equation}
and that on each patch there is a local equivalence map
\begin{equation}
\Phi_\alpha:
U_{2,\alpha}\longrightarrow U_{1,\alpha}\subset\F_1
\label{eq:classical_local_maps_patchwise}
\end{equation}
such that
\begin{equation}
S_2=\Phi_\alpha^\ast S_1
\qquad
\text{on }U_{2,\alpha}.
\label{eq:classical_patchwise_action_equivalence}
\end{equation}
If the probe class includes response observables, the same patchwise map must
also transport the field-space metric,
\begin{equation}
G_2=\Phi_\alpha^\ast G_1
\qquad
\text{on }U_{2,\alpha}.
\label{eq:classical_patchwise_metric_equivalence}
\end{equation}
The question is whether the patchwise data define an equivalence on the larger
domain \(U_2\), rather than only on each patch separately.

The first obstruction is the gluing of the forward map. On an overlap
\begin{equation}
U_{2,\alpha\beta}
=
U_{2,\alpha}\cap U_{2,\beta},
\end{equation}
both \(\Phi_\alpha\) and \(\Phi_\beta\) are defined. In order to obtain a
single-valued map
\begin{equation}
\Phi:U_2\longrightarrow U_1\subset\F_1,
\end{equation}
the local maps must agree on overlaps,
\begin{equation}
\Phi_\alpha
=
\Phi_\beta
\qquad
\text{on }U_{2,\alpha\beta},
\label{eq:classical_forward_gluing}
\end{equation}
up to the relevant gauge identifications when gauge redundancies are present.
If Eq.~\eqref{eq:classical_forward_gluing} fails, then the same configuration
\(\varphi\in U_{2,\alpha\beta}\) is assigned two different images in
\(\F_1\). The patchwise maps then do not define a single field-space map at all.

Equivalently, one may introduce the transition map
\begin{equation}
C_{\alpha\beta}
=
\Phi_\beta^{-1}\circ\Phi_\alpha
\end{equation}
where it is defined. The gluing condition is the requirement that this
transition be trivial,
\begin{equation}
C_{\alpha\beta}
=
\mathrm{id}
\qquad
\text{on }U_{2,\alpha\beta},
\label{eq:classical_forward_transition_trivial}
\end{equation}
again modulo gauge identifications if the endpoint field spaces are quotients.
This is the field-space analogue of the usual condition that local coordinate
identifications must have compatible transition functions in order to define a
single identification on a larger domain.

Even if the forward maps glue, this is not yet enough for an equivalence on the
glued domains. One must also check that the resulting map is invertible there.
Let
\begin{equation}
V_{1,\alpha}
=
\Phi_\alpha(U_{2,\alpha})
\subset\F_1
\end{equation}
be the image patches, and define the local inverse branches
\begin{equation}
\Psi_\alpha
=
\Phi_\alpha^{-1}:
V_{1,\alpha}\longrightarrow U_{2,\alpha}.
\end{equation}
On overlaps of the image patches,
\begin{equation}
V_{1,\alpha\beta}
=
V_{1,\alpha}\cap V_{1,\beta},
\end{equation}
the inverse branches must agree,
\begin{equation}
\Psi_\alpha
=
\Psi_\beta
\qquad
\text{on }V_{1,\alpha\beta}.
\label{eq:classical_inverse_gluing}
\end{equation}
If this condition fails, the forward description may be single-valued while the
inverse is multivalued. The relation is then only branchwise. This is precisely
what happens for a non-monotonic field redefinition, where each branch gives a
valid local map, but overlapping images lead to more than one inverse branch.

When both the forward and inverse gluing conditions hold, the local maps define
an equivalence on the glued domains
\begin{equation}
U_2\subset\F_2,
\qquad
U_1=\Phi(U_2)\subset\F_1.
\end{equation}
These domains need not be the full configuration spaces. It is therefore useful
to define the maximal equivalence domains
\begin{equation}
U_2^{\max}\subset\F_2,
\qquad
U_1^{\max}\subset\F_1
\end{equation}
as the largest domains on which the local maps can be glued into a single
invertible equivalence map
\begin{equation}
\Phi^{\max}:U_2^{\max}\longrightarrow U_1^{\max}
\end{equation}
satisfying
\begin{equation}
S_2=(\Phi^{\max})^\ast S_1
\qquad
\text{on }U_2^{\max},
\end{equation}
and, if response probes are included,
\begin{equation}
G_2=(\Phi^{\max})^\ast G_1
\qquad
\text{on }U_2^{\max}.
\end{equation}
The domains are maximal in the sense that they cannot be enlarged without
violating invertibility, the action-level equivalence, the metric compatibility
required by the probe class, or the gluing conditions on overlaps.

Thus one should distinguish three cases. Local equivalence means that the
criteria of Sec.~\ref{sec:direct_local_classical_equivalence} hold on a single
domain or patch. Branchwise equivalence means that they hold on a family of
patches, but the local maps or their inverses do not glue into a single
invertible map on the union of those patches. Equivalence on maximal domains
means that the local data have been glued as far as possible, giving an
invertible map
\begin{equation}
\Phi^{\max}:U_2^{\max}\longrightarrow U_1^{\max}.
\end{equation}
Global direct equivalence is the special case in which the maximal domains are
the full configuration spaces,
\begin{equation}
U_2^{\max}=\F_2,
\qquad
U_1^{\max}=\F_1.
\end{equation}

The same distinction applies to parent-mediated equivalence. Suppose that an
endpoint domain \(U_i\subset\F_i\) is covered by patches
\begin{equation}
U_i=\bigcup_\alpha U_{i,\alpha},
\end{equation}
and that on each patch there is a reduction section
\begin{equation}
s_{i,\alpha}:U_{i,\alpha}\longrightarrow \widetilde\F.
\end{equation}
These local sections define a single reduction leg
\begin{equation}
s_i:U_i\longrightarrow\widetilde\F
\end{equation}
only if they agree on overlaps,
\begin{equation}
s_{i,\alpha}
=
s_{i,\beta}
\qquad
\text{on }U_{i,\alpha}\cap U_{i,\beta},
\label{eq:classical_parent_section_gluing}
\end{equation}
or agree modulo the relevant parent gauge redundancy when the parent space is a
gauge-completing extension. If this fails, the endpoint is recovered from the
parent only patchwise.

For scalar observables, gluing of the parent sections ensures that the
reduced actions and reduced observables define single-valued endpoint objects:
\begin{equation}
S_i=s_{i,\alpha}^{\ast}\widetilde S,
\qquad
\OO_D^{(i)}=s_{i,\alpha}^{\ast}\widetilde\OO_D
\qquad
\text{on }U_{i,\alpha},
\end{equation}
with compatible values on overlaps. If response probes are included, the
endpoint metrics induced from the parent must glue as well:
\begin{equation}
G_i=s_{i,\alpha}^{\ast}\widetilde G
\qquad
\text{on }U_{i,\alpha},
\label{eq:classical_parent_metric_patchwise}
\end{equation}
with
\begin{equation}
s_{i,\alpha}^{\ast}\widetilde G
=
s_{i,\beta}^{\ast}\widetilde G
\qquad
\text{on }U_{i,\alpha}\cap U_{i,\beta}.
\label{eq:classical_parent_metric_gluing}
\end{equation}
Otherwise the parent construction may define compatible endpoint actions but
not compatible endpoint response tensors.

The parent case can also fail to extend beyond a branch because the reduction
equation has several branches. For example, the equations used to eliminate
auxiliary fields may define local solutions
\begin{equation}
\Xi_{\star,\alpha}(\Phi_{\rm end})
\end{equation}
on different patches, but no single-valued solution
\begin{equation}
\Xi_\star:\F_{\rm end}\longrightarrow\F_\Xi
\end{equation}
on the full endpoint domain. Then the corresponding sections
\begin{equation}
s_\alpha(\Phi_{\rm end})
=
(\Phi_{\rm end},\Xi_{\star,\alpha}(\Phi_{\rm end}))
\end{equation}
define a branchwise parent reduction, not a single reduction on the union. This
is the reduction analogue of the inverse-branch obstruction in the direct case.

As in the direct case, one may define maximal parent-mediated equivalence
domains
\begin{equation}
U_i^{\max}\subset\F_i,
\qquad
i=1,2,
\end{equation}
as the largest endpoint domains on which the local reduction sections,
parent-space reparametrizations and, when required, induced metrics glue into a
single parent-mediated chain. Parent-mediated global equivalence is the special
case in which these maximal endpoint domains are the full endpoint
configuration spaces,
\begin{equation}
U_i^{\max}=\F_i,
\qquad
i=1,2.
\end{equation}

In summary, equivalence is not a binary property of the actions alone. It depends on the domain, on the chosen branch of the
field-space map or reduction, on the probe class being tested, and on whether
the local data glue to a single relation.

\section{Quantum equivalence}
\label{sec:quantum_equiv_section}

The classical criteria of Section~\ref{sec:classical_equiv_section} identify when
two descriptions represent the same classical theory relative to a chosen probe
class. This is still not a criterion for quantum equivalence. Quantization requires
additional choices, such as the functional measure, gauge fixing, the source prescription in the generating functional and the renormalization scheme. Even if two classical representations are related by a direct map or by a parent-mediated chain, independently quantizing the two endpoints need not
produce equivalent quantum theories.

The issue is especially sharp for parent constructions. Classically, a parent
leg may be implemented by solving auxiliary equations or by restricting to a
section of the enlarged configuration space. Quantum mechanically, reducing and quantizing need not commute. Starting from an endpoint
representation, one may quantize directly and obtain an endpoint effective
action. Alternatively, one may quantize the parent representation and only then
perform the corresponding effective reduction. These two procedures need not
give the same result.

Thus classical equivalence supplies the geometrical starting
point, but it does not uniquely determine the quantum object to be compared. The quantum
criterion will therefore be formulated at the effective action level. The central role
will be played by the Vilkovisky--DeWitt effective action introduced in Section~\ref{sec:vdw_section}, whose scalar
transformation law and covariant derivative tensors provide the appropriate
off-shell quantum data.

\subsection{Renormalized effective representations}
\label{sec:renormalized_effective_representations}

We package the output of the quantum construction into the renormalized
Vilkovisky--DeWitt effective action. At a stated
accuracy \(N\), for instance in an EFT expansion, we define a renormalized
effective representation by
\begin{equation}
\MM_{\rm eff}^{(N)}
=
\bigl(
\F,\Gg,G,\GammaVDW_N[\Phi;c]
\bigr).
\label{eq:Meff_quantum_section}
\end{equation}
Here \(\F\) is the configuration space, \(\Gg\) is the gauge group when present,
\(G_{AB}\) is the field-space metric used in the VDW construction, and
\(\GammaVDW_N[\Phi;c]\) is the renormalized VDW effective action expressed in
terms of renormalized parameters \(c^I\). In an EFT, the \(c^I\) include the
Wilson coefficients retained at the chosen order. Related recent discussions
of equivalence at the effective-action level appear in \cite{Cohen:2023wxr,Falls:2025effective}.

The field-space metric appears explicitly in
Eq.~\eqref{eq:Meff_quantum_section} because it is needed to define the VDW
measure, covariant source prescription and covariant derivatives of the effective action. It is therefore part of a full VDW effective
representation, unlike the classical case where the metric is only needed for response observables. Consequently, equivalence
of effective representations requires the geometrical data entering the VDW
construction to be identified by the field-space map.

When comparing two renormalized effective actions, one must also specify how
their renormalized parameters are identified. The same quantum theory may be
written with different finite parametrizations of its couplings or Wilson
coefficients. In particular, if the two effective actions were renormalized using
independent renormalization conditions, their parameters need not have the same
physical meaning. One must either translate both actions to a common
renormalization convention or match both parameter sets to the same physical
input observables. Only after this matching has been specified does it make
sense to ask whether the two effective actions agree.

A finite change of scheme or matching convention changes
coordinates on the renormalized parameter space,
\begin{equation}
c^I\longmapsto c^{\prime I}(c),
\end{equation}
and correspondingly changes the local part of the renormalized effective action.
Such a change is not new physics if it can be absorbed into a redefinition of
the parameters already present in the theory, up to the stated accuracy. By
contrast, if no such parameter map exists, or if the difference requires an
independent local operator not contained in the chosen parametrization, then the
two effective actions describe distinct quantum theories rather than different
renormalization conventions.

We shall write
\begin{equation}
\Gamma_N \simeq_N \Gamma'_N
\label{eq:effective_action_equivalence_accuracy}
\end{equation}
when the two renormalized effective actions agree after finite
reparametrizations of the parameters already present in the theory, up to terms
beyond the stated accuracy \(N\). In a renormalizable theory this means equality
after finite redefinitions of the renormalized couplings. In an EFT it also
allows differences that lie beyond the chosen truncation order.

This is not a quotient by arbitrary local functionals. A local operator with an
independent Wilson coefficient defines a new direction in theory space. It is
not a removable ambiguity unless that direction is already part of the parameter
space being reparametrized, or unless it lies beyond the declared accuracy.
Thus \(\simeq_N\) removes finite renormalization conventions, but it does not erase
genuine local physics.

\subsection{Direct local quantum equivalence}
\label{sec:direct_local_quantum_equivalence}

As in the classical discussion, equivalence is always meant relative to a
chosen probe class \(\mathfrak D\). This idea is straightforwardly extended to the quantum realm, where now the underlying operational requirement is that the corresponding scalar quantum observables agree. Since this condition remains the same as in Eqs.~\eqref{eq:scalarequiv} (for direct equivalence) and \eqref{eq:parent_observable_pullback_classical} (for mediated equivalence), we shall not repeat it again. Instead, we state the effective structures whose transport is sufficient to implement it for the quantum probe classes considered below.

Let
\begin{equation}
\MM_{{\rm eff},1}^{(N)}
=
(\F_1,\Gg_1,G_1,\GammaVDW_{1,N}[\Phi_1;c_1]),
\qquad
\MM_{{\rm eff},2}^{(N)}
=
(\F_2,\Gg_2,G_2,\GammaVDW_{2,N}[\Phi_2;c_2])
\end{equation}
be two renormalized effective representations. In view of the operational
criterion stated above, we shall say that they are directly locally quantum
equivalent relative to the quantum probe class under consideration on \(U_2\subset\F_2\), if there exists a local field-space diffeomorphism
\begin{equation}
\Phi:U_2\longrightarrow U_1\subset\F_1
\label{eq:direct_quantum_map}
\end{equation}
together with an identification of renormalized
parameters,
\begin{equation}
c_1^I=c_1^I(c_2),
\label{eq:quantum_parameter_matching}
\end{equation}
such that
\begin{equation}
\GammaVDW_{2,N}[\varphi;c_2]
\simeq_N
\GammaVDW_{1,N}[\Phi(\varphi);c_1(c_2)]
\qquad
\text{on }U_2 .
\label{eq:direct_quantum_equivalence_parameter_form}
\end{equation}
For gauge theories, the map \eqref{eq:direct_quantum_map} must also preserve
gauge orbits (see Eq.~\eqref{eq:gauge_equivariance_classical_section}).

We note that, since
\(G_{AB}\) is part of the VDW construction, full equivalence of effective
representations also requires the same map to transport the field-space metric,
\begin{equation}
G_2=\Phi^\ast G_1
\qquad
\text{on }U_2 .
\label{eq:direct_quantum_metric_equivalence}
\end{equation}
Thus Eqs.~\eqref{eq:direct_quantum_equivalence_parameter_form} and
\eqref{eq:direct_quantum_metric_equivalence} are the quantum counterparts of
direct local classical equivalence at the level of VDW effective
representations.
If one wishes to compare only a restricted set of scalar quantum observables after
the effective actions have already been constructed, one may use a weaker
criterion involving only the scalar values in that probe class. This comparison obviously does not imply equivalence of VDW effective
representations and does not imply equality of covariant response tensors.

Several consequences of Eq.~\eqref{eq:direct_quantum_equivalence_parameter_form}
are worth mentioning. First, once the effective actions agree as scalar
functionals, the matching of their covariant derivative towers is not an
independent assumption. If the field-space geometries are transported as in
Eq.~\eqref{eq:direct_quantum_metric_equivalence}, then differentiating
Eq.~\eqref{eq:direct_quantum_equivalence_parameter_form} covariantly gives
\begin{equation}
(\GammaVDW_{2,N})_{;A_1\cdots A_n}(\varphi;c_2)
\simeq_N
\frac{\delta\Phi^{B_1}}{\delta\varphi^{A_1}}
\cdots
\frac{\delta\Phi^{B_n}}{\delta\varphi^{A_n}}
(\GammaVDW_{1,N})_{;B_1\cdots B_n}
(\Phi(\varphi);c_1(c_2))
\qquad
\text{on }U_2 .
\label{eq:covariant_vdw_vertex_tower_quantum_direct}
\end{equation}
Thus the equality of covariant vertices follows from the scalar equality of the
VDW effective actions, together with the transport of the field-space geometry.
One should not impose vertex matching as an additional condition. It is already
contained in the quantum equivalence criterion.
It then follows that scalar observables built from these tensors and from
consistently transported probes agree. For instance, if a response observable is
locally represented by
\begin{equation}
\OO^{(n)}_{D}[\Phi;c]
=
(\GammaVDW_N)_{;A_1\cdots A_n}[\Phi;c]\,
D^{A_1\cdots A_n},
\end{equation}
then Eq.~\eqref{eq:covariant_vdw_vertex_tower_quantum_direct} implies
\begin{equation}
\OO^{(n)}_{2,D}[\varphi;c_2]
\simeq_N
\OO^{(n)}_{1,\Phi_\ast D}[\Phi(\varphi);c_1(c_2)]
\label{eq:quantum_observable_equivalence}
\end{equation}
for every probe \(D\) in the chosen probe class. If the probe class is a
separating test class, equality of all such pairings is equivalent to equality
of the corresponding covariant theory tensors. If one restricts instead to a
smaller physical probe class, the resulting equivalence is weaker but more
operational.

Second, local tensorial diagnostics extracted from the covariant Hessian are
preserved under an admissible field redefinition. Define
\begin{equation}
H_{AB}[\Phi;c]
=
(\GammaVDW_N)_{;AB}[\Phi;c] .
\label{eq:vdw_covariant_hessian_quantum_direct}
\end{equation}
Under Eq.~\eqref{eq:direct_quantum_equivalence_parameter_form} and
Eq.~\eqref{eq:direct_quantum_metric_equivalence}, the Hessians are related by
pullback,
\begin{equation}
(H_2)_{AB}(\varphi;c_2)
\simeq_N
\frac{\delta\Phi^C}{\delta\varphi^A}
\frac{\delta\Phi^D}{\delta\varphi^B}
(H_1)_{CD}(\Phi(\varphi);c_1(c_2)) .
\label{eq:vdw_hessian_pullback_quantum_direct}
\end{equation}
Equivalently, after raising one index with the transported field-space metric, namely
\begin{equation}
\mathcal H^A{}_B
=
G^{AC}H_{CB}
\ ,
\label{eq:mixed_vdw_hessian_quantum_direct}
\end{equation}
the fluctuation operators are related by a similarity transformation. Therefore,
the local perturbative content around a background is preserved. In
particular, within a fixed choice of boundary conditions, a field redefinition cannot change the number of physical fluctuation
modes, create or remove zero modes, or alter the constrained versus propagating
character of the perturbations.
This gives a useful necessary diagnostic for equivalence, for if two proposed
descriptions have different local perturbative
degree-of-freedom content around a background in \(U_2\), then they cannot be related by a field redefinition on that domain.

Like in the classical case, the strength of this equivalence statement depends on the probe class. If the class is a
separating test class $\mathfrak D_{\rm test}$, equality of all such pairings is equivalent to equality
of the underlying effective theory tensors, up to the stated accuracy. If one restricts instead to a smaller
physical probe class \(\mathfrak D_{\rm phys}\), the resulting notion of
equivalence is weaker but more operational. This distinction is what allows the
same formalism to include both the standard on-shell use of field redefinitions
and the stronger off-shell comparisons needed in semiclassical gravity,
cosmology and non-equilibrium QFT.

\subsection{Parent-mediated quantum equivalence}
\label{sec:parent_mediated_quantum_equivalence}

In the classical parent construction, the endpoint theory is recovered only after imposing the
auxiliary equations of motion that define a chosen reduction leg. Thus the classical
equivalence only holds along the auxiliary solution subspace, not in the full enlarged configuration space. This distinction becomes essential for quantization.

Let
\begin{equation}
\MM_{{\rm cl},{\rm end}}
=
(\F_{\rm end},\Gg_{\rm end},S_{\rm end},G_{\rm end})
\label{eq:classical_endpoint_rep_quantum_section}
\end{equation}
be a classical endpoint representation, and let
\begin{equation}
\widetilde\MM_{\rm cl}
=
(\widetilde\F,\widetilde\Gg,\widetilde S,\widetilde G)
\label{eq:classical_parent_rep_quantum_section}
\end{equation}
be a parent representation. Relative to the endpoint $\MM_{{\rm cl},{\rm end}}$, write
the parent fields as
\begin{equation}
\widetilde\Phi=(\Phi,\Xi),
\end{equation}
where \(\Phi\in\F_{\rm end}\) are the endpoint fields and \(\Xi\) denotes the
auxiliary variables. Classically, the reduction is specified by
auxiliary equations (see Section~\ref{sec:parent_mediated_classical_equivalence})
\begin{equation}
E_{\Xi,\alpha}[\Phi,\Xi]
\equiv
\frac{\delta\widetilde S}{\delta\Xi}
=0,
\label{eq:parent_auxiliary_equations_quantum_section}
\end{equation}
which, on a chosen nondegenerate branch, determine
\begin{equation}
\Xi=\Xi_\star(\Phi)
\quad \text{on }U_{\rm end} \subset \mathcal F_{\rm end}
.
\end{equation}
Equivalently, they define a section
\begin{equation}
s:\F_{\rm end}\longrightarrow\widetilde\F,
\qquad
s(\Phi)=(\Phi,\Xi_\star(\Phi)),
\label{eq:quantum_section_classical_parent_leg}
\end{equation}
such that
\begin{equation}
S_{\rm end}
=
s^\ast\widetilde S .
\label{eq:classical_endpoint_parent_action_reduction_quantum_section}
\end{equation}
For response probes, the corresponding field-space geometries must also be
compatible, namely
\begin{equation}
G_{\rm end}=s^\ast\widetilde G
\ .
\label{eq:classical_endpoint_parent_metric_reduction_quantum_section}
\end{equation}

The question now is whether the classical parent-mediated equivalence survives
quantization in the operational sense stated above. Since the quantum criterion
is implemented at the effective-action level, this becomes the question whether
the endpoint effective action obtained by direct quantization agrees with the
endpoint effective action induced by the parent construction.
Starting from
\(\MM_{{\rm cl},{\rm end}}\), with its own measure and source prescription as in Eqs.~\eqref{eq:covariant_generating_functional_vdw}, \eqref{eq:W_definition_vdw} and \eqref{eq:GammaVDW_from_W}, one obtains
\begin{equation}
\Gamma_{{\rm end},N}^{\rm direct}[\Phi;c_{\rm end}] ,
\label{eq:direct_endpoint_effective_action_quantum_section}
\end{equation}
where \(c_{\rm end}\) denotes the renormalized parameters.
This is the VDW effective action of the endpoint theory itself, computed to the
stated accuracy \(N\).
The quantization of the parent $\widetilde\MM_{\rm cl}$, on the other hand, gives the same quantum theory \eqref{eq:direct_endpoint_effective_action_quantum_section} only if the parent path
integral implements the same reduction condition \eqref{eq:parent_auxiliary_equations_quantum_section}. If the parent variables were integrated freely, the path integral would
sample configurations away from the surface \(E_\Xi=0\). That is a legitimate
quantization of the enlarged theory, but it is not automatically the
quantization of the endpoint theory. More importantly, the condition $E_\Xi=0$ is algebraic, hence it is a constraint and not a dynamical equation.

Therefore, the correct quantization of the parent representation is performed as a constrained functional integral. Locally, on a nondegenerate branch $\Xi_\star(\Phi)$, the constrained path integral is implemented via the delta functional $\delta[E_\Xi]$:
\begin{equation}
\int \D\Xi\,
\delta[E_\Xi]\,
\Delta_\Xi\,
F[\Phi,\Xi]
=
F[\Phi,\Xi_\star(\Phi)]
\ ,
\label{eq:auxiliary_delta_reduction_identity}
\end{equation}
for an arbitrary functional $F[\Phi,\Xi]$ and Jacobian
\begin{equation}
\Delta_\Xi[\Phi,\Xi]
=
\left|
\Det
\left(
\frac{\delta E_{\Xi,\alpha}[\Phi,\Xi]}
{\delta \Xi^\beta}
\right)
\right|
\ .
\label{eq:auxiliary_reduction_jacobian}
\end{equation}
This projects the integration domain on the reduction surface $E_\Xi=0$.
From Eq.~\eqref{eq:auxiliary_delta_reduction_identity}, we can then define the parent generating functional:
\begin{equation}
\exp\left\{
\frac{i}{\hbar}W_{\rm end}^{\rm parent}[J]
\right\}
=
\int
\D\Phi\,\D\Xi\,
\widetilde\mu[\Phi,\Xi] \,
\delta[E_\Xi]\,
\Delta_\Xi[\Phi,\Xi]\,
\exp\left\{
\frac{i}{\hbar}
\left[
\widetilde S[\Phi,\Xi]
+
J_A\,\widetilde\sigma_{\rm end}^A
\right]
\right\},
\label{eq:endpoint_adapted_parent_quantization}
\end{equation}
with measure
\begin{equation}
    \widetilde\mu[\Phi,\Xi] = \sqrt{\Det \widetilde G[\Phi,\Xi]}
    \ .
    \label{eq:measure}
\end{equation}
Note that in this construction the source \(J_A\) is coupled only to the endpoint variables, whereas the variables
\(\Xi\) remain sourceless.

From Eqs.~\eqref{eq:auxiliary_delta_reduction_identity} and \eqref{eq:endpoint_adapted_parent_quantization}, one finds
\begin{equation}
\exp\left\{
\frac{i}{\hbar}W_{\rm end}^{\rm parent}[J]
\right\}
=
\int
\D\Phi\,
\widetilde\mu[\Phi,\Xi_\star(\Phi)]\,
\exp\left\{
\frac{i}{\hbar}
\left[
\widetilde S[\Phi, \Xi_\star(\Phi)]
+
J_A\,\sigma_{\rm end}^A
\right]
\right\}.
\label{eq:parent_generator_reduced_to_endpoint}
\end{equation}
The reduced density appearing in Eq.~\eqref{eq:parent_generator_reduced_to_endpoint}
is
\begin{equation}
\mu_{\rm red}[\Phi]
\equiv
(s^\ast\widetilde\mu)[\Phi]
=
\widetilde\mu[\Phi,\Xi_\star(\Phi)] .
\label{eq:parent_reduced_density_quantum_section}
\end{equation}
Here \(s^\ast\widetilde\mu\) denotes the density obtained by restricting the
parent density to the reduction section in the split coordinates
\((\Phi,\Xi)\). This should not be confused with the Riemannian density obtained
from the pulled-back metric. In general,
\begin{equation}
s^\ast\widetilde\mu
=
\sqrt{\Det\widetilde G[\Phi,\Xi_\star(\Phi)]}
\neq
\sqrt{\Det(s^\ast\widetilde G)} .
\label{eq:measure_pullback_vs_pulledback_metric}
\end{equation}
The difference is the contribution of the auxiliary directions to the full
parent determinant. Therefore, if the direct endpoint theory is defined with
the VDW measure
\begin{equation}
\mu_{\rm end}[\Phi]
=
\sqrt{\Det G_{\rm end}[\Phi]},
\end{equation}
the constrained parent quantization agrees with it only if
\begin{equation}
\mu_{\rm red}=\mu_{\rm end}.
\label{eq:parent_measure_compatibility_quantum_section}
\end{equation}

Eq.~\eqref{eq:parent_measure_compatibility_quantum_section} is a measure-compatibility condition, not a consequence of
\(G_{\rm end}=s^\ast\widetilde G\) alone.
However, since the parent metric is an extension of the endpoint metric, its auxiliary
components are part of the chosen parent data. One may therefore choose an
endpoint-adapted extension for which the reduced density $\mu_{\rm red}$ agrees with the
endpoint VDW density $\mu_{\rm end}$. To see what this condition requires, let us work locally in adapted coordinates to the surface $\Xi_\star(\Phi)$ by shifting the auxiliary variables:
\begin{equation}
\eta^\alpha
=
\Xi^\alpha-\Xi^\alpha_\star(\Phi).
\end{equation}
In the variables \((\Phi,\eta)\), the reduction surface is simply
\(\eta^\alpha=0\), and variations of the endpoint fields are tangent to this
surface.
On this surface, and after imposing \(G_{\rm end}=s^\ast\widetilde G\), the
parent metric can be written in block form as
\begin{equation}
\widetilde G\big|_{\eta=0}
=
\begin{pmatrix}
G_{\rm end} & K\\
K^T & L
\end{pmatrix},
\label{eq:endpoint_adapted_parent_metric_extension}
\end{equation}
where \(K\) contains the mixed endpoint--auxiliary components and \(L\) is the
auxiliary block. The determinant then factorizes as
\begin{equation}
\Det\widetilde G\big|_{\eta=0}
=
\Det G_{\rm end}\,
\Det\!\left(
L-K^T G_{\rm end}^{-1}K
\right).
\end{equation}
Therefore the measure-compatibility condition
\(\mu_{\rm red}=\mu_{\rm end}\) is equivalent, up to field-independent
normalization, to
\begin{equation}
\Det\!\left(
L-K^T G_{\rm end}^{-1}K
\right)
=
\text{constant}.
\label{eq:endpoint_adapted_normal_determinant}
\end{equation}
This is the general endpoint-adaptedness condition on the auxiliary part of the
parent metric. A simple sufficient choice is \(K=0\) and
\(L=\mathbf 1_{\rm aux}\) on the reduction surface. In particular, the usual tacit assumption of flat metrics automatically satisfies the measure compatibility condition \eqref{eq:parent_measure_compatibility_quantum_section}.

With such an
endpoint-adapted choice, the constrained parent integral uses the same action,
source prescription and measure as the direct endpoint integral. The action reduction
\eqref{eq:classical_endpoint_parent_action_reduction_quantum_section}, metric
compatibility \eqref{eq:classical_endpoint_parent_metric_reduction_quantum_section}
and measure compatibility \eqref{eq:parent_measure_compatibility_quantum_section}
then give
\begin{equation}
W_{\rm end}^{\rm parent}[J]
=
W_{\rm end}^{\rm direct}[J]
\quad
\text{on }U_{\rm end}
,
\label{eq:equal_endpoint_generators}
\end{equation}
hence the corresponding effective actions agree:
\begin{equation}
\Gamma_{\rm end}^{\rm parent}[\Phi; c_{\rm end}]
=
\Gamma_{\rm end}^{\rm direct}[\Phi; c_{\rm end}]
\quad
\text{on }U_{\rm end}
\ .
\end{equation}
If the two computations use different renormalization conventions, this
equality should be replaced by equality after parameter matching,
\begin{equation}
{\GammaVDW}_{{\rm end},N}^{\rm direct}[\Phi;c_{\rm end}]
\simeq_N
{\GammaVDW}_{{\rm end},N}^{\rm parent}
[\Phi;\widetilde c_{\rm end}(c_{\rm end})]
\qquad
\text{on }U_{\rm end}.
\label{eq:equal_endpoint_effective_actions}
\end{equation}
The point is not that the constrained parent integral defines a new quantum
theory, but that it identifies the precise quantum object equivalent to the
endpoint theory. Freely quantizing the enlarged field space is a different
operation.
We also stress that, whilst in the classical case the transport of the metric is required only for response functions, at the quantum level the metric must always be transported due to the functional measure \eqref{eq:measure} and the covariant source coupling.

The delta functional can also be implemented by introducing Lagrange
multipliers $\Lambda^\alpha$:
\begin{equation}
\delta[E_\Xi]
=
\int \D\Lambda\,
\exp\left\{
\frac{i}{\hbar}
\Lambda^\alpha E_{\Xi,\alpha}
\right\},
\label{eq:delta_exponentiation_parent}
\end{equation}
with the determinant in Eq.~\eqref{eq:endpoint_adapted_parent_quantization}
kept as part of the reduced measure or represented by the corresponding ghosts.
Thus a nonlinear reduction condition may be imposed either directly by the
delta functional (as in Eqs.~\eqref{eq:endpoint_adapted_parent_quantization}--\eqref{eq:parent_generator_reduced_to_endpoint}) or by enlarging the parent once more with multiplier fields. In the latter case, it becomes clear that the bare action
\begin{equation}
    S_{\rm aux}[\Phi,\Xi,\Lambda]
    \equiv
    \widetilde{S}[\Phi,\Xi] + \Lambda^\alpha E_{\Xi,\alpha}
    \label{eq:auxaction}
\end{equation}
leads to the same quantum theory as $\widetilde{S}[\Phi,\Xi]$. In particular,
when the auxiliary variable is already a Lagrange multiplier, the delta
functional is generated automatically upon integration over it. This is the case of $f(R)$, as we shall see in Section~\ref{sec:examples_section}.

One may also quantize the enlarged parent theory as a theory in its own right,
with sources for all parent mean fields and without constraints. This gives a full parent VDW effective
action
\begin{equation}
\widetilde\GammaVDW_N[\Phi,\Xi;\widetilde c].
\label{eq:full_parent_effective_action_quantum_section}
\end{equation}
Reducing this object by the quantum equation
\begin{equation}
\frac{\delta\widetilde\GammaVDW_N}{\delta\Xi^\alpha}=0
\label{eq:full_parent_quantum_reduction_equation}
\end{equation}
is an effective-action-level analogue of classical elimination. This is a
legitimate construction for the enlarged quantum theory, but it should not be
confused with the endpoint-adapted constrained quantization in Eq.~\eqref{eq:endpoint_adapted_parent_quantization}. In particular, beyond the tree level, the direct and parent effective actions obtained via Eqs.~\eqref{eq:full_parent_effective_action_quantum_section}--\eqref{eq:full_parent_quantum_reduction_equation} will not satisfy Eq.~\eqref{eq:equal_endpoint_effective_actions}. Eliminating the constraint $E_\Xi=0$ clearly modifies the theory.

\subsection{Branchwise quantization}
\label{sec:branchwise_quantization}

The discussion above was formulated on a chosen nondegenerate branch of the
field-space map or reduction section. This is the natural setting for local
equivalence. When the map is not globally one-to-one, the original quantum
theory can still be rewritten in the new variables, but not as a single path
integral over one branch. The correct rewriting is a sum over all
branches, with the appropriate Jacobian on each branch. This branch-summed
object is equivalent to the original path integral. It should not be confused
with the quantum theory obtained by selecting only one endpoint branch.
Throughout this discussion, possible contributions from degenerate critical
sets on which the functional Jacobian vanishes are excluded; they require a
separate analysis and are not covered by the local or branchwise equivalence
criterion.

Consider first an ordinary change of variables. Let
\begin{equation}
\phi=F(\chi)
\label{eq:branchwise_field_map}
\end{equation}
be a field-space map that is locally invertible but not globally one-to-one.
For a fixed value of \(\phi\), the equation
\begin{equation}
\phi=F(\chi)
\end{equation}
may have several regular solutions,
\begin{equation}
\chi=\chi_r(\phi),
\qquad
\Det\left(
\frac{\delta F}{\delta\chi}
\right)_{\chi=\chi_r(\phi)}
\neq0 ,
\label{eq:inverse_branches_direct_quantum}
\end{equation}
where \(r\) labels the inverse branch. The variable \(\phi\) alone is then not
a faithful global coordinate, because the same value of \(\phi\) can come from
different values of \(\chi\). To rewrite the original path integral without
losing information, one must keep track of the branch label \(r\).
Thus the transformed description is not a single theory written only in terms
of \(\phi\). It rather comprises many sectors labelled by the inverse branches
\(\chi_r(\phi)\).

This can be seen directly from the path integral. Starting from
\begin{equation}
Z_\chi[J]
=
\int \D\chi\,
\sqrt{\Det G_\chi[\chi]}\,
\exp\left\{
\frac{i}{\hbar}
\left[
S_\chi[\chi]
+
J_A\sigma_\chi^A(\bar\chi,\chi)
\right]
\right\},
\label{eq:Z_chi_branchwise_start}
\end{equation}
insert the identity
\begin{equation}
1
=
\int \D\phi\,
\delta[\phi-F(\chi)] .
\label{eq:branch_identity_insert}
\end{equation}
For each regular inverse branch \(\chi=\chi_r(\phi)\), let \(V_r\) denote the
domain of \(\phi\)-configurations on which that branch exists and remains
nondegenerate. The delta functional then gives
\begin{equation}
\int \D\chi\,
\delta[\phi-F(\chi)]\,A[\chi]
=
\sum_r
A[\chi_r(\phi)]
\left|
\Det
\left(
\frac{\delta F}{\delta\chi}
\right)_{\chi=\chi_r(\phi)}
\right|^{-1},
\label{eq:functional_delta_sum_over_roots}
\end{equation}
where the sum is over the regular branches.

The original generating functional can then be rewritten as
\begin{equation}
Z_\chi[J]
=
\sum_r
\int_{V_r}
\D\phi\,
\sqrt{\Det G_{\phi,r}[\phi]}\,
\exp\left\{
\frac{i}{\hbar}
\left[
S_r[\phi]
+
J_A\sigma_{\chi,r}^A(\bar\chi,\phi)
\right]
\right\}
\ ,
\label{eq:Z_branch_sum_direct}
\end{equation}
where we defined
\begin{equation}
S_r[\phi]
\equiv
S_\chi[\chi_r(\phi)]
\label{eq:branch_action_direct}
\end{equation}
and
\begin{equation}
\sigma_{\chi,r}^A(\bar\chi,\phi)
\equiv
\sigma_\chi^A(\bar\chi,\chi_r(\phi)).
\label{eq:branch_source_pullback_direct}
\end{equation}
The metric \(G_{\phi,r}\) is the metric induced on the \(r\)-th branch,
\begin{equation}
G_{\phi,r}
=
\chi_r^\ast G_\chi .
\label{eq:branch_metric_direct}
\end{equation}
Indeed, its Riemannian measure is
\begin{equation}
\sqrt{\Det G_{\phi,r}[\phi]}
=
\sqrt{\Det G_\chi[\chi_r(\phi)]}
\left|
\Det
\left(
\frac{\delta F}{\delta\chi}
\right)_{\chi=\chi_r(\phi)}
\right|^{-1}.
\label{eq:branch_riemannian_measure_direct}
\end{equation}
Thus the Jacobian appearing from the delta functional is precisely the Jacobian
needed to rewrite the original Riemannian measure in the branch variables.

Therefore, the full quantization in the original variable is equal to a sum over
branch contributions,
\begin{equation}
Z_\chi[J]
=
\sum_r Z_r[J],
\label{eq:Z_equals_sum_branch_Z}
\end{equation}
where \(Z_r[J]\) denotes the \(r\)-th term in
Eq.~\eqref{eq:Z_branch_sum_direct}. A single branch then defines a
local or branchwise quantum representation. The full original quantization is
recovered only when all regular branches are included, with the correct domains
and branch measures. Finally, we note that
\begin{equation}
W_\chi[J]\neq\sum_r W_r[J]
\ ,
\end{equation}
hence the VDW effective action of the full branch-summed theory is not,
in general, the sum of the branch effective actions.

The same logic applies to parent reductions. Suppose that, for a fixed endpoint
configuration \(\Phi\), the auxiliary equations
\begin{equation}
E_{\Xi,\alpha}[\Phi,\Xi]=0
\end{equation}
have several regular solutions
\begin{equation}
\Xi=\Xi_r(\Phi),
\qquad
s_r(\Phi)=(\Phi,\Xi_r(\Phi)).
\label{eq:multiple_parent_sections_quantum}
\end{equation}
Let \(U_r\subset\F_{\rm end}\) denote the endpoint domain on which the branch
\(s_r\) exists and remains nondegenerate. Then, with the same determinant
factor used in the constrained parent integral,
\begin{equation}
\int \D\Xi\,
\delta[E_\Xi]\,
\Delta_\Xi\,
F[\Phi,\Xi]
=
\sum_r
F[\Phi,\Xi_r(\Phi)] .
\label{eq:parent_delta_sum_over_branches}
\end{equation}
Thus the constrained parent generator \eqref{eq:endpoint_adapted_parent_quantization} decomposes into branch
contributions:
\begin{equation}
Z_{\rm end}^{\rm parent}[J]
=
\sum_r
Z_{{\rm end},r}^{\rm parent}[J],
\label{eq:parent_Z_sum_over_branches}
\end{equation}
where
\begin{equation}
Z_{{\rm end},r}^{\rm parent}[J]
=
\int_{U_r}
\D\Phi\,
\sqrt{\Det G_{{\rm end},r}[\Phi]}\,
\exp\left\{
\frac{i}{\hbar}
\left[
S_{{\rm end},r}[\Phi]
+
J_A\sigma_{{\rm end},r}^A(\bar\Phi,\Phi)
\right]
\right\}.
\label{eq:parent_branch_generator}
\end{equation}
Here
\begin{equation}
S_{{\rm end},r}
=
s_r^\ast\widetilde S,
\qquad
G_{{\rm end},r}
=
s_r^\ast\widetilde G ,
\label{eq:branchwise_parent_induced_data}
\end{equation}
and the parent metric has been chosen in the same endpoint-adapted way on this
branch. Thus the reduced parent measure is the measure of the branch metric,
\begin{equation}
s_r^\ast\widetilde\mu
=
\sqrt{\Det G_{{\rm end},r}}.
\label{eq:branchwise_parent_measure_compatibility}
\end{equation}
Again, this is automatic for flat product metrics in adapted coordinates. The
object \(\sigma_{{\rm end},r}^A\) is the VDW geodesic displacement associated
with the branch metric \(G_{{\rm end},r}\).

Eqs~\eqref{eq:Z_equals_sum_branch_Z} and \eqref{eq:parent_branch_generator} clarify the meaning of branchwise equivalence. If one selects a
single branch \(r\), then one obtains a single endpoint theory on \(U_r\).
Equivalence to the directly quantized endpoint on that branch requires
\begin{equation}
{\GammaVDW}_{{\rm end},N}^{\rm direct}[\Phi;c_{\rm end}]
\simeq_N
{\GammaVDW}_{{\rm end},r,N}^{\rm parent}
[\Phi;\widetilde c_{{\rm end},r}(c_{\rm end})]
\qquad
\text{on }U_r .
\label{eq:branchwise_parent_lift_condition}
\end{equation}
By contrast, if one integrates over all branches, the result is a quantum theory with several sectors.
Such a branch-summed theory is equivalent to a direct endpoint quantization only
if the endpoint theory itself is defined as a sum over the same sectors.

A branchwise description means restricting to a
sector in which the branch choice is fixed and the reduction remains
nondegenerate. Crossings between branches require a separate global analysis and
are not captured by a single local equivalence map.
This distinction is important for the interpretation of global equivalence.
A non-invertible field redefinition or a multivalued reduction can still be used
to rewrite a full path integral, but the result is a sum over sectors. This is not the same as a single globally defined field redefinition
between two endpoint configuration spaces. A global equivalence to one endpoint
description requires a single nondegenerate branch.

\subsection{Domains and gluing at the quantum level}
\label{sec:domains_gluing_quantum}

The local, branchwise and global distinctions are the same as in
Sec.~\ref{sec:local_branchwise_global_classical_equivalence}. The difference is
the structure being transported. In the direct classical case the local
condition was
\begin{equation}
S_2=\Phi^\ast S_1,
\end{equation}
with \(G_2=\Phi^\ast G_1\) when response probes were included. In the quantum
case this is replaced by
\begin{equation}
\GammaVDW_{2,N}
\simeq_N
\Phi^\ast\GammaVDW_{1,N} .
\label{eq:quantum_gluing_replacement_direct}
\end{equation}

Thus a patchwise family of quantum equivalences
\begin{equation}
\Phi_\alpha:
U_{2,\alpha}\longrightarrow U_{1,\alpha}
\end{equation}
must satisfy the same forward- and inverse-gluing requirements as before. The
maps must glue to a single field-space map on the domain under consideration,
and the inverse branches must glue on overlaps of the image patches. If the
forward maps fail to glue, there is no single field-space map. If the forward
map glues but the inverse branches do not, the equivalence is only branchwise.

The only additional bookkeeping is that the equality is an equality of
renormalized effective actions in the sense of \(\simeq_N\). Different patches
may use different renormalized parametrizations, but on overlaps these
parametrizations must describe the same effective theory after parameter
matching. A transition from one patch to another may therefore change the finite
coordinates \(c^I\) on parameter space, but it must not introduce an independent
operator direction within the stated accuracy.

The maximal quantum equivalence domains
\begin{equation}
U_2^{\max}\subset\F_2,
\qquad
U_1^{\max}\subset\F_1
\end{equation}
are the largest domains on which the local maps and their inverse branches glue,
and on which
\begin{equation}
\GammaVDW_{2,N}
\simeq_N
(\Phi^{\max})^\ast\GammaVDW_{1,N}
\end{equation}
holds. Quantum global equivalence is the special case in which these maximal
domains are the full configuration spaces being compared,
\begin{equation}
U_2^{\max}=\F_2,
\qquad
U_1^{\max}=\F_1 .
\end{equation}
If the theories are formulated from the start on selected sectors
\(\F_i^{(0)}\subset\F_i\), for example fixed-signature,
fixed-boundary-condition or fixed-branch sectors, then in practice global equivalence means
\begin{equation}
U_i^{\max}=\F_i^{(0)},
\qquad
i=1,2.
\end{equation}

For parent-mediated relations, the same statement applies with the replacement
\begin{equation}
S_i=s_i^\ast\widetilde S
\qquad
\longrightarrow
\qquad
{\GammaVDW}_{i,N}^{\rm direct}
\simeq_N
{\GammaVDW}_{i,N}^{\rm parent}.
\end{equation}
The maximal domain \(U_i^{\rm max}\subset\F_i\)
is the largest endpoint domain on which the parent-induced endpoint effective
action is well defined and satisfies
\begin{equation}
{\GammaVDW}_{i,N}^{\rm direct}
\simeq_N
{\GammaVDW}_{i,N}^{\rm parent}.
\end{equation}
In the constrained construction described above, the same local data that define
\({\GammaVDW}_{i,N}^{\rm parent}\) also determine the domain of the classical
section and the induced metric. Thus the quantum domain is
as local as the underlying parent reduction. Global parent-mediated quantum equivalence holds when this maximal domain is
the whole endpoint configuration spaces, i.e. $U_i^{\rm max}=\F_i$.

\subsection{Theory-space interpretation}
\label{sec:theory_space_interpretation}

We can also interpret the above equivalence criteria using equivalence classes. Fix an
accuracy \(N\), a probe class \(\mathfrak D\), and a domain prescription
(local, maximal-domain or global). Let \(\mathcal M_{\rm eff}^{(N)}\) denote the
set of all effective representations,
\begin{equation}
\MM_{\rm eff}^{(N)}
=
\bigl(
\F,\Gg,G,\GammaVDW_N[\Phi;c]
\bigr)
\ ,
\end{equation}
admitted at this accuracy.
Different points of \(\mathcal M_{\rm eff}^{(N)}\) may correspond to different
field coordinates, different parent presentations, different choices of
renormalized parameters, or genuinely different quantum theories.

The equivalence criteria above define an equivalence relation on \(\mathcal M_{\rm eff}^{(N)}\). Indeed, let
\(\MM_{{\rm 1},\rm eff}^{(N)}\) and \(\MM_{{\rm 2},\rm eff}^{(N)}\) be two effective representations in
\(\mathcal M_{\rm eff}^{(N)}\). If they satisfy either the direct equivalence condition,
Eq.~\eqref{eq:direct_quantum_equivalence_parameter_form}, or the parent-mediated one,
Eq.~\eqref{eq:equal_endpoint_effective_actions}, then we define
\begin{equation}
\MM_{\rm 1,eff}^{(N)}
\sim_{\mathfrak D}^{(N)}
\MM_{\rm 2,eff}^{(N)}
\ .
\end{equation}
An effective theory, in the sense used here, is then an equivalence class
\begin{equation}
[\MM_{\rm eff}^{(N)}]_{\mathfrak D}
=
\left\{
\MM_{\rm eff}^{\prime (N)}
\in
\mathcal M_{\rm eff}^{(N)}
\ \bigg|\ 
\MM_{\rm eff}^{\prime (N)}
\sim_{\mathfrak D}^{(N)}
\MM_{\rm eff}^{(N)}
\right\}.
\label{eq:effective_theory_equivalence_class}
\end{equation}
The corresponding theory space is the quotient
\begin{equation}
\mathcal T_{\mathfrak D}^{(N)}
=
\mathcal M_{\rm eff}^{(N)}
/
\sim_{\mathfrak D}^{(N)} .
\label{eq:theory_space_quantum_section}
\end{equation}
A point in \(\mathcal M_{\rm eff}^{(N)}\) is therefore a particular
representation, while a point in \(\mathcal T_{\mathfrak D}^{(N)}\) is the
theory described by all representations in the same equivalence class, relative
to the chosen probe class and accuracy. This quotient \eqref{eq:theory_space_quantum_section} should be understood operationally, not ontologically. It
identifies representations that give the same scalar observables for the chosen
probe class, rather than extracting a representation-independent common core.
The structures needed to
define the observables, such as the field-space metric and the probe representatives, remain part of the effective representations
being compared.\footnote{
This operational use of the quotient is related to, but narrower than, recent
philosophical discussions of dualities, common-core realism and
observer-dependent ontology \cite{Dawid:2025knu}. Here our quotient
only classifies effective representations by equality of probe-dependent
observables.
}

This formalizes the standard intuition that distinct actions, field variables,
frames, auxiliary-field presentations and renormalization conventions may
describe the same physics. Field redefinitions and parent-mediated maps (when the parent theory is properly quantized via the constrained path integral) act inside an equivalence class. 
Other transformations can however move one to a different class.

Finally, the equivalences considered here should be distinguished from more
general dualities. A duality need not be induced by a field-space diffeomorphism, nor by
an auxiliary-variable parent reduction. It may instead be given by a dictionary
between quite distinct theories. Holographic
dualities provide the most prominent example. In AdS/CFT, the bulk and boundary
descriptions do not have the same configuration space, they are not related by a parent extension/reduction, and the holographic
relation is not a local change of field variables. Rather, the dictionary
identifies boundary sources with asymptotic bulk data, schematically
\begin{equation}
Z_{\rm bulk}[\phi_{(0)}]
=
Z_{\rm CFT}[J],
\qquad
\phi_{(0)}\leftrightarrow J ,
\end{equation}
after imposing the appropriate boundary conditions and renormalization
prescription. From the viewpoint adopted here, such a relation would be a
generalized equivalence defined directly at the level of observables and probes.
The present paper does not attempt to formulate this more general notion. It rather
develops the narrower, but technically necessary, case in which the dictionary is
generated by field redefinitions or by parent-mediated reductions.

\section{Examples and applications}
\label{sec:examples_section}

\subsection{Scalar EFT: local, branchwise and global direct equivalence}
\label{sec:scalar_application_direct_equivalence}

We first consider a scalar EFT example. The point is not the dynamics of the
model itself, but the distinction between a change of field coordinates and a
genuine change of theory. Since there is no gauge symmetry, the example isolates
the role of the field-space metric, the VDW source prescription, the EFT
accuracy and the domain of the field redefinition.

Let the first classical representation be
\begin{equation}
\MM_{{\rm cl},1}
=
(\F_\phi,\{e\},S_1,G_1),
\end{equation}
where \(\{e\}\) denotes the trivial gauge group. The action is
\begin{equation}
S_1[\phi]
=
\int \dd^dx
\left[
\frac{1}{2}(\partial\phi)^2
-\frac{1}{2}m^2\phi^2
-\frac{\lambda}{4!}\phi^4
\right],
\label{eq:S1_scalar_application}
\end{equation}
and the field-space metric is the flat ultralocal metric
\begin{equation}
(G_1)_{xy}(\phi)
=
\delta^{(d)}(x-y).
\label{eq:G1_scalar_application}
\end{equation}

Now introduce the pointwise field redefinition\footnote{For notational simplicity, we shall adopt the same symbol \(F\) for the pointwise transformation $\phi=F(\varphi)$ and the induced configuration-space map \((F(\varphi))(x)=F(\varphi(x))\).}
\begin{equation}
\phi=F(\varphi)
=
\varphi+\frac{a}{\Lambda^2}\varphi^3 .
\label{eq:scalar_field_redefinition_application}
\end{equation}
Its functional Jacobian is ultralocal,
\begin{equation}
\frac{\delta F(\varphi(x))}{\delta\varphi(y)}
=
F'(\varphi(x))\delta^{(d)}(x-y),
\qquad
F'(\varphi)
=
1+\frac{3a}{\Lambda^2}\varphi^2 .
\label{eq:scalar_jacobian_application}
\end{equation}
Thus \(F\) defines a local configuration-space diffeomorphism
\begin{equation}
F:U\subset\F_\varphi\longrightarrow F(U)\subset\F_\phi
\end{equation}
on any domain \(U\) such that
\begin{equation}
F'(\varphi(x))\neq0
\qquad
\text{for all }x
\text{ and all }\varphi\in U .
\label{eq:scalar_local_invertibility_condition}
\end{equation}

Consider another classical representation 
\begin{equation}
	\MM_{{\rm cl},2}
=
(\F_\varphi,\{e\},S_2,G_2),
\end{equation}
defined by:
\begin{equation}
S_2=F^\ast S_1,
\qquad
G_2=F^\ast G_1 ,
\label{eq:scalar_classical_pullback_data}
\end{equation}
on $U$.
More explicitly, the metric components are
\begin{equation}
(G_2)_{xy}(\varphi)
=
F'(\varphi(x))^2\delta^{(d)}(x-y) ,
\label{eq:G2_scalar_application}
\end{equation}
and the action is
\begin{equation}
S_2[\varphi]
=
S_1[F(\varphi)]
.
\label{eq:scalar_action_pullback_application}
\end{equation}
Expanding to first order in \(\Lambda^{-2}\), we find:
\begin{equation}
S_2[\varphi]
=
S_1[\varphi]
+
\frac{a}{\Lambda^2}
\int \dd^dx\,
\varphi^3(x)
\frac{\delta S_1}{\delta\varphi(x)}
+
O(\Lambda^{-4}).
\label{eq:scalar_eom_expansion_application}
\end{equation}
Since
\begin{equation}
\frac{\delta S_1}{\delta\varphi}
=
-\Box\varphi
-m^2\varphi
-\frac{\lambda}{6}\varphi^3,
\end{equation}
and assuming boundary terms vanish on the class of configurations considered,
this gives
\begin{equation}
S_2[\varphi]
=
S_1[\varphi]
+
\frac{a}{\Lambda^2}
\int \dd^dx
\left[
3\varphi^2(\partial\varphi)^2
-m^2\varphi^4
-\frac{\lambda}{6}\varphi^6
\right]
+
O(\Lambda^{-4}).
\label{eq:S2_scalar_expanded_application}
\end{equation}
Therefore, by construction, the conditions for direct local classical equivalence relative to \(\mathfrak D_{\rm resp}^{\rm cl}\) between $\MM_{{\rm cl},1}$ and $\MM_{{\rm cl},2}$ are satisfied
on \(U\). We note that the equation-of-motion operator in Eq.~\eqref{eq:scalar_eom_expansion_application} appears only after expanding the exact
pullback relation \eqref{eq:scalar_action_pullback_application}. Equivalence is an off-shell criterion, hence it does not require the second term in Eq.~\eqref{eq:scalar_eom_expansion_application} to vanish.

We now lift the comparison to the quantum level. The corresponding effective
representations are
\begin{equation}
\MM_{{\rm eff},1}^{(N)}
=
(\F_\phi,\{e\},G_1,\GammaVDW_{1,N}[\phi;c_1]),
\qquad
\MM_{{\rm eff},2}^{(N)}
=
(\F_\varphi,\{e\},G_2,\GammaVDW_{2,N}[\varphi;c_2]).
\label{eq:scalar_effective_representations_application}
\end{equation}
At one loop, the effective action reads:
\begin{equation}
{\GammaVDW}_i^{(1)}[\Phi_i]
=
S_i[\Phi_i]
+
\frac{i\hbar}{2}
\Tr\ln
\left[
(G_i)^{AC}(S_i)_{;CB}
\right],
\qquad
i=1,2,
\label{eq:scalar_one_loop_vdw_general_application}
\end{equation}
up to counterterms and finite renormalization conventions. In the
\(\phi\)-coordinate the metric is flat, so the covariant Hessian is the
ordinary Hessian:
\begin{equation}
(S_1)_{;xy}
=
\frac{\delta^2 S_1}{\delta\phi(x)\delta\phi(y)}
=
\left[
-\Box_x-m^2-\frac{\lambda}{2}\phi(x)^2
\right]
\delta^{(d)}(x-y).
\label{eq:S1_hessian_scalar_application}
\end{equation}
In the \(\varphi\)-coordinate the metric is field dependent. Its Levi-Civita
connection is given by:
\begin{equation}
\Gamma^{z}_{xy}[G_2]
=
\frac{F''(\varphi(x))}{F'(\varphi(x))}
\delta^{(d)}(x-y)\delta^{(d)}(x-z).
\label{eq:scalar_connection_application}
\end{equation}
Since
\begin{equation}
\frac{\delta S_2}{\delta\varphi(x)}
=
F'(\varphi(x))
\frac{\delta S_1}{\delta\phi(x)}
\bigg|_{\phi=F(\varphi)},
\end{equation}
one obtains
\begin{align}
(S_2)_{;xy}
&=
\frac{\delta^2 S_2}{\delta\varphi(x)\delta\varphi(y)}
-
\Gamma^{z}_{xy}[G_2]
\frac{\delta S_2}{\delta\varphi(z)}
\nonumber\\
&=
F'(\varphi(x))F'(\varphi(y))
(S_1)_{;xy}\big|_{\phi=F(\varphi)}
+
O(\Lambda^{-4}).
\label{eq:S2_covariant_hessian_pullback_application}
\end{align}
The non-tensorial term proportional to
\(F''\,\delta S_1/\delta\phi\) cancels against the connection term. This is the
elementary mechanism by which the VDW construction removes the off-shell
parametrization dependence.

Raising one index with the inverse metric \(G_2^{-1}\), the mixed fluctuation
operator in the \(\varphi\)-description is
\begin{equation}
(G_2)^{xz}(S_2)_{;zy}
=
\frac{1}{F'(\varphi(x))}
\,
(G_1)^{xz}(S_1)_{;zy}\big|_{\phi=F(\varphi)}
\,
F'(\varphi(y))
+
O(\Lambda^{-4}).
\label{eq:scalar_mixed_operator_similarity_application}
\end{equation}
Thus the mixed operators are related by a similarity transformation, up to
terms beyond the retained EFT order. Their functional determinants therefore
agree modulo the regulator choice and the finite local renormalizations already
accounted for by \(\simeq_N\). Combining this with
Eq.~\eqref{eq:scalar_action_pullback_application}, one obtains
\begin{equation}
\GammaVDW_{2,N}[\varphi;c_2]
\simeq_N
\GammaVDW_{1,N}[F(\varphi);c_1(c_2)]
\qquad
\text{on }U .
\label{eq:scalar_quantum_equivalence_result_application}
\end{equation}
Hence the scalar EFT is directly equivalent at the quantum level as well.

It remains to specify the domains where this equivalence holds. If \(a>0\), then
\begin{equation}
F'(\varphi)>0
\end{equation}
for all \(\varphi\), so the field redefinition \eqref{eq:scalar_field_redefinition_application} is globally monotonic. In that case one
may take \(U=\F_\varphi\), and the same field redefinition gives a global
equivalence, provided the global configuration-space sector and boundary
conditions are transported consistently.

If \(a<0\), write \(b=|a|\). The derivative vanishes at
\begin{equation}
\varphi=\pm\varphi_c,
\qquad
\varphi_c=\frac{\Lambda}{\sqrt{3b}} .
\label{eq:scalar_critical_points_application}
\end{equation}
Defining the critical set
\begin{equation}
\mathcal C
=
\left\{
\varphi\in\F_\varphi
\ \big|\ 
F'(\varphi(x))=0
\text{ for some }x
\right\},
\label{eq:scalar_critical_set_application}
\end{equation}
the map can still be used branchwise on \(\F_\varphi\setminus\mathcal C\). For
example, one may take
\begin{align}
U_0
&=
\left\{
\varphi\in\F_\varphi
\ \big|\ 
-\varphi_c<\varphi(x)<\varphi_c
\ \text{for all }x
\right\},
\label{eq:scalar_U0_application}
\\
U_+
&=
\left\{
\varphi\in\F_\varphi
\ \big|\ 
\varphi(x)>\varphi_c
\ \text{for all }x
\right\},
\label{eq:scalar_Uplus_application}
\\
U_-
&=
\left\{
\varphi\in\F_\varphi
\ \big|\ 
\varphi(x)<-\varphi_c
\ \text{for all }x
\right\}.
\label{eq:scalar_Uminus_application}
\end{align}
On each of these domains, the restricted map
\begin{equation}
F_i=F|_{U_i}:U_i\longrightarrow F(U_i)\subset\F_\phi
\end{equation}
is a local diffeomorphism. Therefore the direct classical and quantum
equivalence criteria apply on each branch.
However, the inverse branches on the images do not glue into a single inverse
map. Indeed, the images of the branch domains are
\begin{align}
F(U_-)
&=
\left\{
\phi\in\F_\phi
\ \bigg|\
\phi(x)>-\frac{2}{3}\varphi_c
\ \text{for all }x
\right\},
\\
F(U_0)
&=
\left\{
\phi\in\F_\phi
\ \bigg|\
-\frac{2}{3}\varphi_c<\phi(x)<\frac{2}{3}\varphi_c
\ \text{for all }x
\right\},
\\
F(U_+)
&=
\left\{
\phi\in\F_\phi
\ \bigg|\
\phi(x)<\frac{2}{3}\varphi_c
\ \text{for all }x
\right\}.
\end{align}
Their common overlap is therefore
\begin{equation}
V_{\rm ov}
=
F(U_-)\cap F(U_0)\cap F(U_+)
=
\left\{
\phi\in\F_\phi
\ \bigg|\
-\frac{2}{3}\varphi_c<\phi(x)<\frac{2}{3}\varphi_c
\ \text{for all }x
\right\}.
\end{equation}
For every \(\phi\in V_{\rm ov}\), the equation
\begin{equation}
\phi(x)=F(\varphi(x))
\end{equation}
has three different roots, one in each branch. Thus the three local inverses
\begin{equation}
F_-^{-1},\qquad F_0^{-1},\qquad F_+^{-1}
\end{equation}
are all defined on \(V_{\rm ov}\), but they assign different
\(\varphi\)-configurations to the same \(\phi\)-configuration. Hence they
cannot be restrictions of a single inverse map. The cubic field redefinition consequently
defines branchwise equivalences for \(a<0\), but not a single global
equivalence on the full \(\varphi\)-configuration space.

This conclusion concerns the exact cubic completion
\eqref{eq:scalar_field_redefinition_application}. At the level of the truncated
EFT, the situation is slightly subtler. The map \(F\) has only been specified
through order \(\Lambda^{-2}\). One may choose a different exact completion
that agrees with it to this order but is globally invertible. For example,
consider
\begin{equation}
\widetilde F(\varphi)
=
\varphi
+\frac{a}{\Lambda^2}\varphi^3
+\frac{c}{\Lambda^4}\varphi^5,
\qquad
a<0 ,
\label{eq:scalar_global_completion_application}
\end{equation}
with \(c>0\). Then
\begin{equation}
\widetilde F'(\varphi)
=
1+\frac{3a}{\Lambda^2}\varphi^2+\frac{5c}{\Lambda^4}\varphi^4 .
\label{eq:scalar_global_completion_derivative_application}
\end{equation}
Writing \(y=\varphi^2/\Lambda^2\ge0\), the derivative is
\begin{equation}
\widetilde F'(\varphi)
=
1+3ay+5cy^2 .
\end{equation}
For \(a<0\), its minimum occurs at
\begin{equation}
y_\ast=-\frac{3a}{10c},
\end{equation}
and has value
\begin{equation}
\widetilde F'_{\rm min}
=
1-\frac{9a^2}{20c}.
\end{equation}
Thus, if
\begin{equation}
c>\frac{9a^2}{20},
\label{eq:scalar_global_completion_condition_application}
\end{equation}
then \(\widetilde F'(\varphi)>0\) for all \(\varphi\), thus $\widetilde F$ is injective. Since
\(\widetilde F(\varphi)\to\pm\infty\) as \(\varphi\to\pm\infty\), $\widetilde F$ is also surjective and this
pointwise map defines a global diffeomorphism on $\mathbb R$. It therefore induces a global configuration-space diffeomorphism:
\begin{equation}
\widetilde F:\F_\varphi \longrightarrow \F_\phi .
\end{equation}
Moreover,
\begin{equation}
\widetilde F(\varphi)
=
F(\varphi)+O(\Lambda^{-4}),
\end{equation}
so it induces the same EFT pullback data through the retained order. Therefore
the failure of the particular cubic map to globalize for \(a<0\) is not, by
itself, a proof that the truncated EFT representations are globally inequivalent.
At finite EFT order, global equivalence is therefore not determined by the
truncated local field redefinition alone. It is a statement about a specified
configuration-space sector and a chosen exact representative, or about an
explicitly defined class of admissible global completions.

This example exhibits the basic logic of the formalism in the simplest setting.
Classical equivalence on \(\mathfrak D_{\rm kin}^{\rm cl}\) follows from the
pullback of the action. Equivalence on \(\mathfrak D_{\rm resp}^{\rm cl}\) also
requires the pullback of the field-space metric. Direct quantum equivalence for
the covariant vertices is implemented by the equality
\eqref{eq:scalar_quantum_equivalence_result_application} of the renormalized
VDW effective actions.
Local, branchwise and global equivalence are then determined by the domains on
which the field-space map is invertible and by whether the corresponding
branches glue.

\subsection{Quantum frame equivalence and mismatched quantizations}
\label{sec:fr_quantum_mismatched_objects_application}

The classical relations between metric \(f(R)\) gravity and its different representations were already used in
Section~\ref{sec:lessons_classical_equivalence} to motivate the formalism. The
purpose here is to clarify a common source of claims of quantum inequivalence
in the frame and scalar--tensor literature \cite{Kamenshchik:2014waa,Ruf:2017xon,Ohta:2017fqs,Falls:2018utl,Finn:2019aip}.

As discussed in Section~\ref{sec:parent_mediated_quantum_equivalence}, the same classical
relation can be quantized in different ways. One may quantize the metric
endpoint directly, impose the parent reduction inside the path integral, or
instead quantize the enlarged scalar--tensor field space as an independent
theory. These are not the same quantum construction. Thus a disagreement between
the resulting effective actions does not, by itself, show that the quantum
equivalence has failed. It may simply show that different quantum objects have
been compared.

In the case of \(f(R)\), this mismatch can occur in two distinct places. The
first is the parent step, where one must distinguish an endpoint-adapted
reduction from the unconstrained quantization of an enlarged theory. The second
is the Jordan--Einstein step, where the map is an ordinary field redefinition on
its domain, but ordinary off-shell effective actions are not field-space
scalars.

Let us first consider the parent step. As discussed in
Section~\ref{sec:parent_mediated_quantum_equivalence}, a parent representation
does not reproduce an endpoint quantum theory merely because the corresponding
classical actions agree after eliminating auxiliary variables. The parent path
integral must implement the same reduction. In the \(f(R)\) case this point is
particularly transparent, because the parent action already contains a
Lagrange multiplier:
\begin{equation}
S_\lambda[g,\chi,\lambda]
=
\frac{\Mp^2}{2}\int \dd^4x\,\sqrt{-g}\,
\big(
f(\chi)+\lambda(R-\chi)
\big).
\label{eq:fr_Slambda_quantum_application}
\end{equation}
If this parent is used to describe the metric \(f(R)\) endpoint, then
\(\chi\) and \(\lambda\) are auxiliary fields and they must not be coupled to the source $J$.
The source is
coupled only to the endpoint variable \(g\), and the multiplier imposes the
metric reduction. Suppressing overall normalization factors, the
\(\lambda\)-dependent part of the path integral gives:
\begin{equation}
\int \D_g\lambda\,
\exp\left\{
\frac{i}{\hbar}
\frac{\Mp^2}{2}
\int \dd^4x\,\sqrt{-g}\,
\lambda(R-\chi)
\right\}
\propto
\delta_g\!\left[R-\chi\right],
\label{eq:fr_lambda_generates_delta_application}
\end{equation}
where \(\D_g\lambda\) and \(\delta_g\) denote, respectively, the multiplier
measure and delta functional associated with the geometrical pairing
\(\langle\lambda,C\rangle_g=\int\dd^4x\,\sqrt{-g}\lambda C\) (see Appendix~\ref{app:delta_density_conventions}). Thus the parent integral reduces to
\begin{align}
Z_f^{\rm parent}[J]
&=
\int
\D g\,\D\chi\,
\mu_\lambda[g,\chi]\,
\delta_g\!\left[R-\chi\right]
\nonumber\\
&\hspace{1cm}\times
\exp\left\{
\frac{i}{\hbar}
\left[
\frac{\Mp^2}{2}\int \dd^4x\,\sqrt{-g}\,f(\chi)
+
J_A\widetilde\sigma_f^A
\right]
\right\}
\nonumber\\
&=
\int
\D g\,
(s_f^\ast\mu_\lambda)[g]\,
\exp\left\{
\frac{i}{\hbar}
\left[
S_f[g]
+
J_A\sigma_f^A
\right]
\right\},
\label{eq:fr_parent_reduces_to_metric_generator_application}
\end{align}
where
\begin{equation}
s_f(g)=(g,R[g])
\end{equation}
is the metric endpoint section, $S_f$ is the $f(R)$ action (see Eq.~\eqref{eq:Sf_intro_section}) and parent (resp. endpoint) objects are denoted with (resp. without) the tilde. The parent construction reproduces the direct metric quantization when the source prescription and the measure are inherited by the metric endpoint. In practice this means choosing the parent metric as an endpoint-adapted extension, so that
\begin{equation}
G_f=s_f^\ast G_\lambda,
\qquad
\mu_f=s_f^\ast\mu_\lambda.
\label{eq:fr_parent_metric_condition_application}
\end{equation}
For flat product metrics in coordinates adapted to \(s_f\), the measure equality
is automatic. At the level of renormalized VDW effective actions,
we thus find:
\begin{equation}
{\GammaVDW}_{f,N}^{\rm direct}[g;c_f]
\simeq_N
{\GammaVDW}_{f,N}^{\rm parent}[g;\widetilde c_f(c_f)] .
\label{eq:fr_correct_parent_quantum_comparison_application}
\end{equation}
This is the concrete \(f(R)\) realization of the multiplier construction in
Section~\ref{sec:parent_mediated_quantum_equivalence}. The delta functional
enforcing the endpoint reduction is generated by the Lagrange multiplier
already present in \(S_\lambda\).

This should be contrasted with freely quantizing the standalone two-field
action \(\widehat S[g,\chi]\) of Eq.~\eqref{eq:Shat_intro_section}. As
discussed in Section~\ref{sec:fr_seed}, this action can reproduce the metric
\(f(R)\) action only on the appropriate branch, and only where the reduction is
nondegenerate. Quantum mechanically, however, the more important point is that
free quantization of \(\widehat S[g,\chi]\) defines a theory on the enlarged
field space \(\F_g\times\F_\chi\). Its VDW effective action,
\begin{equation}
\widehat\GammaVDW_N[g,\chi;\widehat c],
\end{equation}
is therefore an effective action for a two-field quantum theory. Eliminating
\(\chi\) afterwards by
\begin{equation}
\frac{\delta\widehat\GammaVDW_N}{\delta\chi}=0
\label{eq:fr_quantum_elimination_chi_application}
\end{equation}
is not the same operation as imposing the metric endpoint reduction inside the
path integral.

This identifies one source of apparent quantum inequivalence. A comparison
between
\begin{equation}
{\GammaVDW}_{f,N}^{\rm direct}[g]
\end{equation}
and
\begin{equation}
\left.
\widehat\GammaVDW_N[g,\chi]
\right|_{\delta\widehat\GammaVDW_N/\delta\chi=0}
\end{equation}
does not compare two representations of the same quantum theory. It compares
the metric endpoint quantization with an enlarged scalar--tensor quantum
theory. A disagreement between these objects does not, by itself, prove quantum
inequivalence of metric \(f(R)\) gravity and its scalar--tensor representation.
It shows that the quantization prescription has changed.

The Jordan--Einstein step is different. Once a scalar--tensor branch has been
chosen, no auxiliary variable is eliminated. The map
\begin{equation}
\hat g_{\mu\nu}=\phi g_{\mu\nu},
\qquad
\sigma=\sqrt{\frac{3}{2}}\,\Mp\ln\phi
\end{equation}
is an ordinary field-space reparametrization on the domain \(\phi>0\). The
issue is therefore not endpoint reduction, but field-space covariance. Ordinary
background-field effective actions are not scalars on configuration space, so a
mismatch between the naive effective action computed with non-transported measures, sources etc, is not an invariant statement of frame inequivalence.
The invariant off-shell comparison is instead the VDW comparison
\begin{equation}
\GammaVDW_{E,N}[\hat g,\sigma;c_E]
\simeq_N
\GammaVDW_{J,N}
[g(\hat g,\sigma),\phi(\sigma);c_J(c_E)]
\qquad
\text{on }\phi>0,
\label{eq:fr_JE_correct_quantum_comparison_application}
\end{equation}
with all the geometrical data
transported by the frame map. If this relation holds, the Jordan and Einstein
descriptions are quantum equivalent relative to the VDW off-shell probe class
on that domain. If it fails because noncovariant quantum objects were compared,
then the failure belongs to the comparison, not to the underlying field
redefinition.

Global issues can also obstruct or obscure the
comparison. In particular, the scalar--tensor description is local on a
nondegenerate branch of the Legendre map
\begin{equation}
\phi=f'(\chi),
\qquad
f''(\chi)\neq0 .
\end{equation}
But if \(f'\) is not globally one-to-one, then a single Jordan-frame action
describes only one branch. Rewriting the full parent path integral in terms of
\(\phi\) requires a sum over the regular inverse branches, with the
corresponding branch measures, as in
Section~\ref{sec:branchwise_quantization}. A disagreement between a single
Jordan or Einstein branch and the full metric-side quantum theory would then
compare a branchwise representation with a branch-summed theory.

\section{Conclusions}
\label{sec:conclusions_section}

Field redefinitions are indispensable in modern quantum field theory. They organize EFT bases,
remove redundant operators, simplify interactions, connect different classical formulations and, in
gravity, relate Jordan, Einstein and higher-derivative descriptions. Yet the standard language used
to discuss them is not fully adequate once one moves beyond on-shell scattering. The equivalence
theorem is an on-shell statement. It guarantees the invariance of scattering amplitudes under
admissible local field redefinitions. It does not, however, supply an off-shell criterion for when two
quantum descriptions define the same theory.

This limitation matters whenever the relevant observables are not exhausted by asymptotic
scattering amplitudes. In gravity, cosmology, and non-equilibrium
quantum field theory one is typically interested in dynamics, whose main observables include response functions and in-in correlators. These are off-shell quantities. The ordinary effective action is not
suited for an intrinsic comparison of such quantities because it does not transform covariantly under
general field redefinitions. This is the physical reason the Vilkovisky--DeWitt effective action plays a
central role in the present paper.

The framework proposed here is based on three simple ideas.
The first is that equivalence should be formulated in terms of observables. In the language used here, observables are scalar quantities
obtained by combining theory data with probe data. This makes explicit that the
content being compared depends on the class of probes under consideration.
On-shell scattering probes lead to the usual equivalence theorem. Off-shell
response probes, in-in correlators and gravitational observables test more
structures.

Secondly, one must distinguish direct field redefinitions from
parent-mediated equivalences. Some reformulations are ordinary changes of
coordinates on a fixed configuration space. Others first enlarge the field space
by introducing auxiliary variables and then recover an endpoint theory by a
reduction. Metric \(f(R)\) gravity and its scalar--tensor form are of the second
type. Treating these two mechanisms as the same operation is one source of
confusion.

Finally, classical equivalence does not automatically lift to
quantum equivalence. A field redefinition or parent reduction tells us how the
classical action is related across descriptions, but a quantum theory also
depends on the measure, the source prescription, the field-space geometry and
the identification of renormalized parameters. If these data are chosen
independently in two classically related formulations, the result may simply be
two different quantum theories. The invariant comparison must therefore be made
after quantization, at the level of the renormalized Vilkovisky--DeWitt
effective actions, with the geometrical data transported consistently and with
finite parameter redefinitions used only to identify equivalent renormalization
conventions. Differences that can be absorbed by such parameter redefinitions
are representational, whereas differences that require new independent operator
directions are genuine quantum inequivalences.

We have also emphasized that local equivalence does not automatically globalize. The gluing of the
forward maps and the gluing of the inverse branches are logically distinct issues. Their failure
provides genuine global obstructions. The scalar EFT example shows this in the simplest possible
setting. The $f(R)$ chain shows the analogous issue in gravity, now intertwined with parent
reductions and domain restrictions such as $f''(\chi)\neq 0$ and $\phi>0$.

The framework is useful in several concrete ways. It gives a precise off-shell
language for frame questions in scalar--tensor gravity and \(f(R)\) theories.
It clarifies why redundant operators may be invisible to scattering amplitudes
while still changing coordinate-dependent off-shell response functions. It also
separates genuine quantum inequivalence from mismatched comparisons. For
example, the direct metric quantization of \(f(R)\), the endpoint-adapted
constrained parent quantization, and the freely quantized enlarged
scalar--tensor theory are different quantum objects. A disagreement between
them is not necessarily a failure of field-redefinition or frame equivalence.

The question of equivalence therefore does not admit a useful general answer at the
level of bare Lagrangians alone. The answer depends on the observables being
tested, on the domain of the field-space map or reduction, and on the quantum
object used for the comparison. At the off-shell quantum level, the appropriate
object is the Vilkovisky--DeWitt effective action, compared after the relevant
geometrical data and renormalized parameters have been matched. Once formulated
this way, many apparent disagreements become easier to interpret. Some are true
failures of equivalence, while others arise because different observables,
different branches or different quantum theories were compared.

\section*{Acknowledgements}

IK is grateful to the National Council for Scientific and Technological Development -- CNPq (Grant Nos. 303283/2022-0, 401567/2023-0 and 200564/2025-0) for partial financial support.

\appendix

\section{Functional deltas and density conventions}
\label{app:delta_density_conventions}

In several parent constructions, a Lagrange multiplier is used to impose a
local constraint. Since the constraint is a field on spacetime, the precise
form of the corresponding functional delta depends on the convention used for
the spacetime density factors. We summarize here the convention used in the
main text.

Let \(C(x)\) be a scalar constraint. For fixed \(g_{\mu\nu}\), define the
geometrical pairing
\begin{equation}
\langle \lambda,C\rangle_g
=
\int \dd^4x\,\sqrt{-g}\,\lambda(x)C(x).
\label{eq:geometrical_pairing_delta_appendix}
\end{equation}
The measure \(\D_g\lambda\) denotes the formal flat measure associated with
this pairing. More explicitly, if \(\{e_n(x)\}\) is an orthonormal basis with
respect to
\begin{equation}
\int \dd^4x\,\sqrt{-g}\,e_n(x)e_m(x)=\delta_{nm},
\end{equation}
and
\begin{equation}
\lambda(x)=\sum_n\lambda_n e_n(x),
\qquad
C(x)=\sum_n C_n e_n(x),
\end{equation}
then
\begin{equation}
\langle \lambda,C\rangle_g=\sum_n\lambda_n C_n,
\qquad
\D_g\lambda=\prod_n \frac{\dd\lambda_n}{2\pi},
\qquad
\delta_g[C]=\prod_n\delta(C_n).
\end{equation}
With these definitions,
\begin{equation}
\delta_g[C]
=
\int \D_g\lambda\,
\exp\left\{
i\int \dd^4x\,\sqrt{-g}\,\lambda C
\right\}.
\label{eq:covariant_delta_fourier_appendix}
\end{equation}

This notation should not be confused with a coordinate product measure
\(\prod_x \dd\lambda(x)\). If one uses such a densitized convention instead,
the same Fourier integral may be written in terms of a delta functional of
\(\sqrt{-g}C\), schematically
\begin{equation}
\delta[\sqrt{-g}\,C]
\sim
\Det(\sqrt{-g})^{-1}\delta[C],
\end{equation}
where the determinant is formal and regulator-dependent. This is not a new
physical factor by itself; it reflects a redistribution of density factors
between the multiplier measure, the delta functional, and the reduced endpoint
measure.

Thus, in the \(f(R)\) parent integral, the invariant content of
\begin{equation}
\int \D_g\lambda\,
\exp\left\{
\frac{i}{\hbar}
\frac{\Mp^2}{2}
\int \dd^4x\,\sqrt{-g}\,\lambda(R-\chi)
\right\}
\end{equation}
is simply that the multiplier imposes
\begin{equation}
R-\chi=0.
\end{equation}
Any convention-dependent normalization associated with this delta functional is
included in the induced endpoint measure.

\bibliographystyle{unsrt}
\bibliography{refs}

\end{document}